\documentclass{article}

\usepackage[main, final]{neurips_2026}

\usepackage[utf8]{inputenc} 
\usepackage[T1]{fontenc}    
\usepackage{hyperref}       
\usepackage{url}            
\usepackage{booktabs}       
\usepackage{amsfonts}       
\usepackage{nicefrac}       
\usepackage{microtype}      
\usepackage{xcolor}         

\usepackage{algorithm}
\usepackage{algorithmic}
\usepackage{multirow}
\usepackage{colortbl}

\usepackage{amssymb}
\usepackage{pifont}
\usepackage{makecell}
\usepackage{tipa}
\usepackage{bbm}
\usepackage{graphicx}
\usepackage{amsmath}
\usepackage{amssymb}
\usepackage{caption}
\usepackage{float}       
\usepackage[table]{xcolor} 
\usepackage{wrapfig}

\definecolor{LightBlue}{rgb}{0.85 0.92 0.96}
\definecolor{custom_gray}{gray}{.92}

\definecolor{darkgreen}{RGB}{0,130,0}
\definecolor{darkred}{RGB}{180,0,0}

\definecolor{cellcol}{gray}{.92}

\title{Seeing Speech: Learning Visible Articulatory Dynamics for Speech-Driven 3D Facial Animation}

\author{%
  Hyung Kyu Kim \\
  Chung-Ang University \\
  Seoul, South Korea \\
  \texttt{hyung1208@cau.ac.kr} \\
  \And
  Byungchan Hwang \\
  Chung-Ang University \\
  Seoul, South Korea \\
  \texttt{byungchan0705@cau.ac.kr} \\
  \And
  Hak Gu Kim\thanks{Corresponding Author} \\
  Chung-Ang University \\
  Seoul, South Korea \\
  \texttt{hakgukim@cau.ac.kr} \\
}

\begin{document}

\maketitle


\begin{abstract}
\label{main:abstract}
Recent progress in speech-driven 3D facial animation has improved vertex-level reconstruction quality, but speech-consistent visible articulation remains difficult. This is because speech production follows structured and constrained articulators' coordination and the mapping from acoustics to motion is inherently one-to-many. Motivated by the structured patterns of visible articulation, we propose a novel articulation-aware framework that models visible speech through directional articulatory motions and composes them into surface-consistent 3D facial motion. To represent visible articulation with three directional articulatory motions, spreading, opening, and protrusion, we propose a Speech--Articulatory Memory (SAM) that captures the correspondence between speech and these motions under phonetic context through retrieval and decoding based on a key-value memory structure.
Then, a Topology-aware Articulatory Composition (TAC) integrates the predicted directional articulatory motions under mesh topology to produce surface-consistent 3D facial motion. Experiments on VOCASET and TFHP show that our method achieves state-of-the-art performance on standard reconstruction metrics and improves visible articulatory distance and velocity errors for lip articulation, while a user study confirms clear preference in lip sync and realism.
\end{abstract}

\section{Introduction}
\label{main:intro}
Speech-driven 3D facial animation aims to generate realistic 3D facial motion from speech.
This enables improved communication in applications such as VR, content production, and education~\cite{tanaka2022acceptability}.
However, generating speech-consistent 3D facial motion is challenging due to one-to-many mapping between acoustics and facial motion~\cite{atal1978inversion, panchapagesan2011study, ghosh2010generalized}.
This stems from articulatory coupling and compensatory mechanisms, where multiple articulator configurations under structured articulatory constraints can produce the same acoustic realization~\cite{maeda1990compensatory, parrell18_interspeech}.
Therefore, modeling these articulatory characteristics is important for generating speech-consistent facial motion.

Despite this, most existing speech-driven facial animation methods do not explicitly model articulatory motion constraints.
Recent approaches, including direct mapping with temporal CNNs~\cite{VOCA2019, CompositeandRegionalFacialMovements}, Transformer-based auto-regressive prediction~\cite{faceformer2022, codetalker2023}, and diffusion-based denoising frameworks~\cite{FaceDiffuser23, diffposetalk2024, facetalk2024, chen2025diffusiontalker, yang2026streamingtalker}, primarily focus on improving reconstruction quality at the vertex or mesh level.
By formulating speech-driven animation as a holistic face-motion regression problem, these methods synthesize overall 3D facial trajectories without explicitly accounting for the structured articulatory motions underlying speech production~\cite{icip2024, kim25r_interspeech}.
As a result, holistic vertex regression offers limited interpretability and weak physical grounding in the mechanisms of speech articulation.

\begin{figure}[t!]
    \centering
    \vspace{-17pt}
    \includegraphics[width=1.0\linewidth]{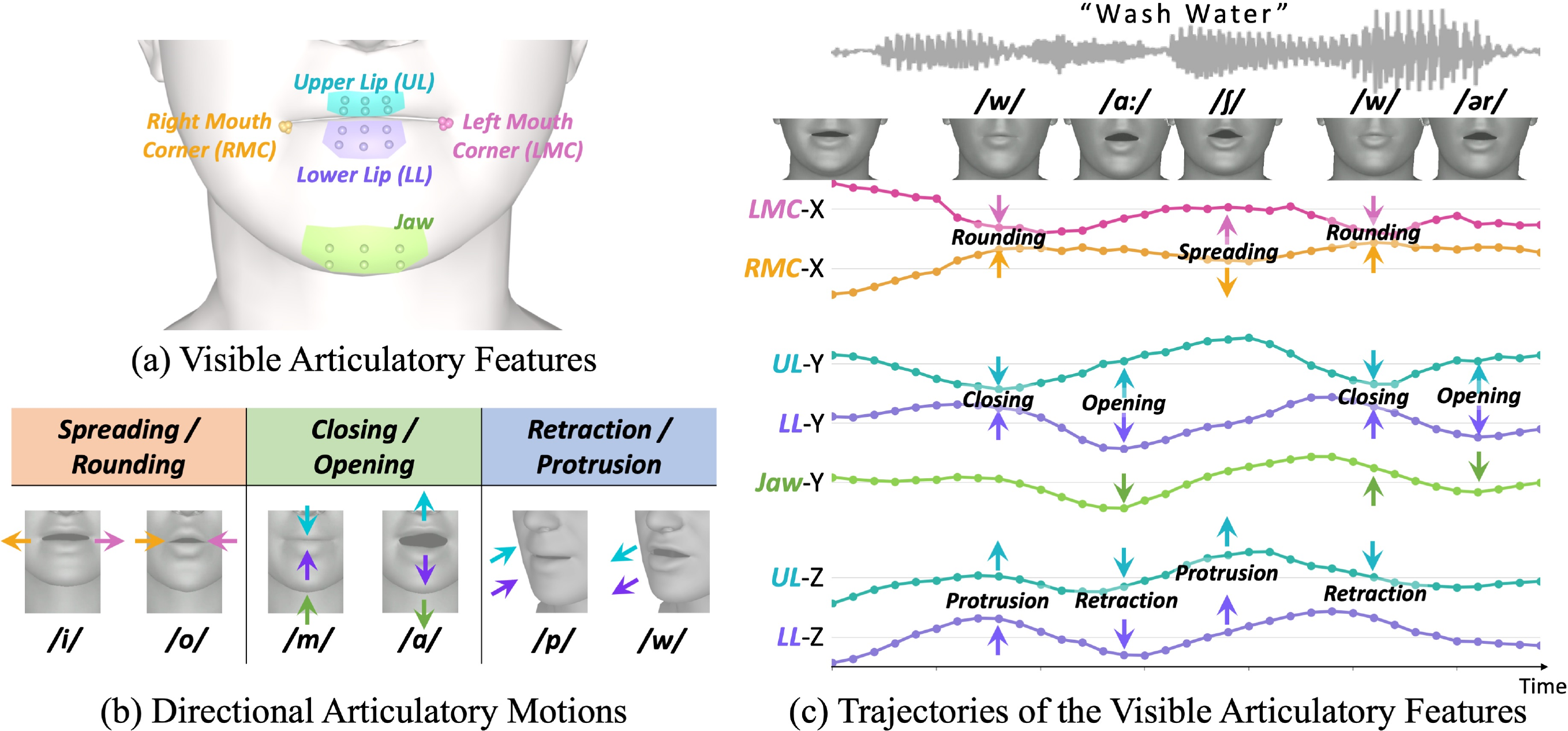}
    \vspace{-12pt}
    \caption{
        How do we model physically grounded and articulatory-consistent facial motion from speech?
        (a) Five visible mesh-surface anchors as motion reference points: upper lip (UL), lower lip (LL), left mouth corner (LMC), right mouth corner (RMC), and Jaw.
        (b) From these anchors, we describe visible articulation with three directional articulatory motions: \textit{horizontal} rounding/spreading, \textit{vertical} opening/closing, and \textit{depth-wise} protrusion/retraction.
        (c) For an utterance, the trajectories of these motions form temporally localized articulatory events that align with acoustic/phonemic cues.
    }
    \vspace{-10pt}
    \label{fig_1_thumb}
\end{figure}

This raises an intriguing research question: \textit{Can we model speech-driven 3D facial animation in a way that is physically grounded in human speech production?}
Human speech is generated through coordinated, time-varying configurations of articulators~\cite{rebernik2021review}.
The lips and jaw are externally observable and play a central role in shaping intelligible speech.
This suggests that speech-consistent facial animation should be grounded in structured visible articulatory motions~\cite{maeda1990compensatory, international1999handbook, anand2025teaching}.
Motivated by this perspective, we examine how \textit{visible articulation} evolves during speech production.
\textit{Visible articulation} refers to the observable movements and configurations of speech organs, particularly the lips and jaw, in producing speech~\cite{icip2024}.
As illustrated in Fig.~\ref{fig_1_thumb} (a) and (b), we define three directional articulatory motions (i.e., Spreading--Rounding, Opening--Closing, and Protrusion--Retraction) based on mesh-surface anchors (UL, LL, LMC, RMC, and Jaw), which specify the directions of lip--jaw articulations~\cite{lofqvist2005lip, lucero2008analysis}.
Across representative phonemes, we observe that visible articulation is not arbitrary, but follows consistent, directionally dominant motion patterns (see Fig.~\ref{fig_1_thumb} (c)). 

In this paper, we propose a novel speech-driven 3D facial animation framework grounded in the structured nature of visible articulation during human speech production.
Based on this observation, we interpret speech-driven facial motion as arising from directionally organized articulatory motions that recur across phonemes and reflect consistent visible articulatory coordination patterns.
For this purpose, we first introduce a Speech--Articulatory Memory (SAM) to capture the correspondence between speech and directional articulatory motions, modeling how speech gives rise to specific patterns of visible articulation under phonetic context.
Second, we propose a Topology-aware Articulatory Composition (TAC) to translate directional articulatory motions into surface-consistent 3D facial motion, ensuring that visible articulation is realized consistently over the facial surface under the mesh topology.
By grounding speech-driven facial animation in structured visible articulation and its directional organization, our approach emphasizes interpretability and physical consistency with human speech production.

\noindent Our main contributions are summarized as follows:
\begin{itemize}
    \item To the best of our knowledge, we propose the first \textit{articulation}-aware framework for speech-driven 3D facial animation grounded in structured visible articulation.    
    \item We propose a novel memory network named Speech--Articulatory Memory (SAM) to model the relationship between speech and directional articulatory motions. SAM captures phonetic context-dependent visible articulation through memory-based retrieval and decoding.
    \item We propose a Topology-aware Articulatory Composition (TAC) to compose directional articulatory motions into surface-consistent 3D facial motion. TAC ensures consistent visible articulation over the facial surface under mesh topology.
    \item With quantitative and qualitative evaluations, including user studies, we demonstrate the effectiveness of SAM and TAC. Our method shows superior performance and improved interpretability over state-of-the-art speech-driven 3D facial animation models.
\end{itemize}

\section{Related Works}
\subsection{Speech-Driven 3D Facial Animation}
Speech-driven 3D facial animation aims to synthesize 3D facial motion from speech~\cite{meshtalk2021, yang2023semi, li2025wav2sem, thambiraja2025diface, he2023speech4mesh, zhuang2024learn2talk, chatziagapi2023avface,CompositeandRegionalFacialMovements, FaceXHuBERT_ICMI23, haque2025probtalk3d, zhuang2026talkingeyes, zhang2026ex}.
Early approaches try to map audio features directly to facial motion using temporal CNNs~\cite{VOCA2019} or Transformers for long-range dependencies~\cite{faceformer2022}. 
Subsequent methods improve motion fidelity with discrete motion priors~\cite{codetalker2023, wu2024probtalk3d, yang2024probabilistic}, key-motion priors~\cite{kmtalk2024}, and self-supervised lip-reading supervision~\cite{selftalk23}.
Beyond motion modeling, UniTalker~\cite{unitalker2024} unifies learning across heterogeneous annotations~\cite{VOCA2019, BIWI2010, emotalk23, wuu2023multifacedatasetneuralface}, and ScanTalk~\cite{scantalk2024} removes fixed-topology constraints through DiffusionNet~\cite{DiffusionNet} on unregistered scans.
For richer animation, recent methods disentangle pose~\cite{chu2025artalkspeechdriven3dhead, diffposetalk2024}, speaker-specific style~\cite{imitator2023, kim2025memorytalker, mimic2024, chu2025dcptalk}, or emotion~\cite{danvevcek2023emotional, emotalk23, xieecoface, pan2025model, kim2025deeptalk, lin2024emoface, qu2025exptalk, shen2024deitalk, jiang2026editemotalk, 10.1145/3757377.3763887, chu2026siatalker, chu2026eetalk} using large-scale in-the-wild mesh data~\cite{diffposetalk2024, emotalk23, mimic2024}.

However, these approaches mainly improve motion fidelity, topology handling, or controllability within audio-to-face generation, but they do not explicitly model the visible articulatory structure underlying speech production.
Prior articulatory studies show that visible speech exhibits lip–jaw motion patterns, including lip opening, horizontal spreading or rounding, and protrusion or retraction~\cite{lofqvist2005lip, lucero2008analysis, chartier2018encoding, wang2013articulatory}.
Motivated by this observation, we represent visible articulation with three directional articulatory motions and compose them over the mesh topology to produce 3D facial motion, grounding generation in the physical mechanisms of speech production~\cite{icip2024, kim25r_interspeech}.

\subsection{Articulatory Modeling for Speech and Animation}
Acoustic-to-Articulatory Inversion (AAI) studies how articulator movements can be inferred from speech acoustics, revealing the structured relationship between sound and human speech production~\cite{sun22b_interspeech, chung24_interspeech}.
Prior studies show that specific phonemes induce dominant, direction-specific motions in visible articulators~\cite{fromkin1964lip, montgomery1983physical}.
Visible speech is characterized by phoneme-dependent and spatially localized motion patterns, including variations in lip aperture, horizontal spreading or rounding, and depth-wise protrusion or retraction~\cite{lofqvist2005lip, lucero2008analysis, wang2013articulatory, georges22_interspeech}.
These components are interpretable from a speech production perspective and predictable from acoustic signals~\cite{craig2008linear}.
Building on these insights, several animation methods incorporate articulatory structure through parameterized representations.
JALI~\cite{edwards2016jali} decomposes facial motion into lip and jaw action parameters~\cite{badin2002three}.
VisemeNet~\cite{zhou2018visemenet} predicts these parameters from audio.
In such approaches, each phoneme is associated with a predefined viseme, and facial motion is generated by blending these fixed shapes.
While this provides a structured control space, articulation is reduced to a low-dimensional parameter set, where context-dependent variation is expressed through combinations of predefined viseme patterns.

However, articulatory motion in natural speech is not expressed as combinations of fixed shapes, but as continuous changes in vertex positions that vary with phonetic context.
To address this limitation, we move beyond viseme-based parameterization and model speech-driven facial motion directly at the vertex level.
Specifically, we represent articulation through three canonical directions and predict how speech drives these direction-specific motions across the facial surface.



\begin{figure*}[t!]
\begin{center}
\vspace{-30pt}
\includegraphics[width=1.05 \linewidth]{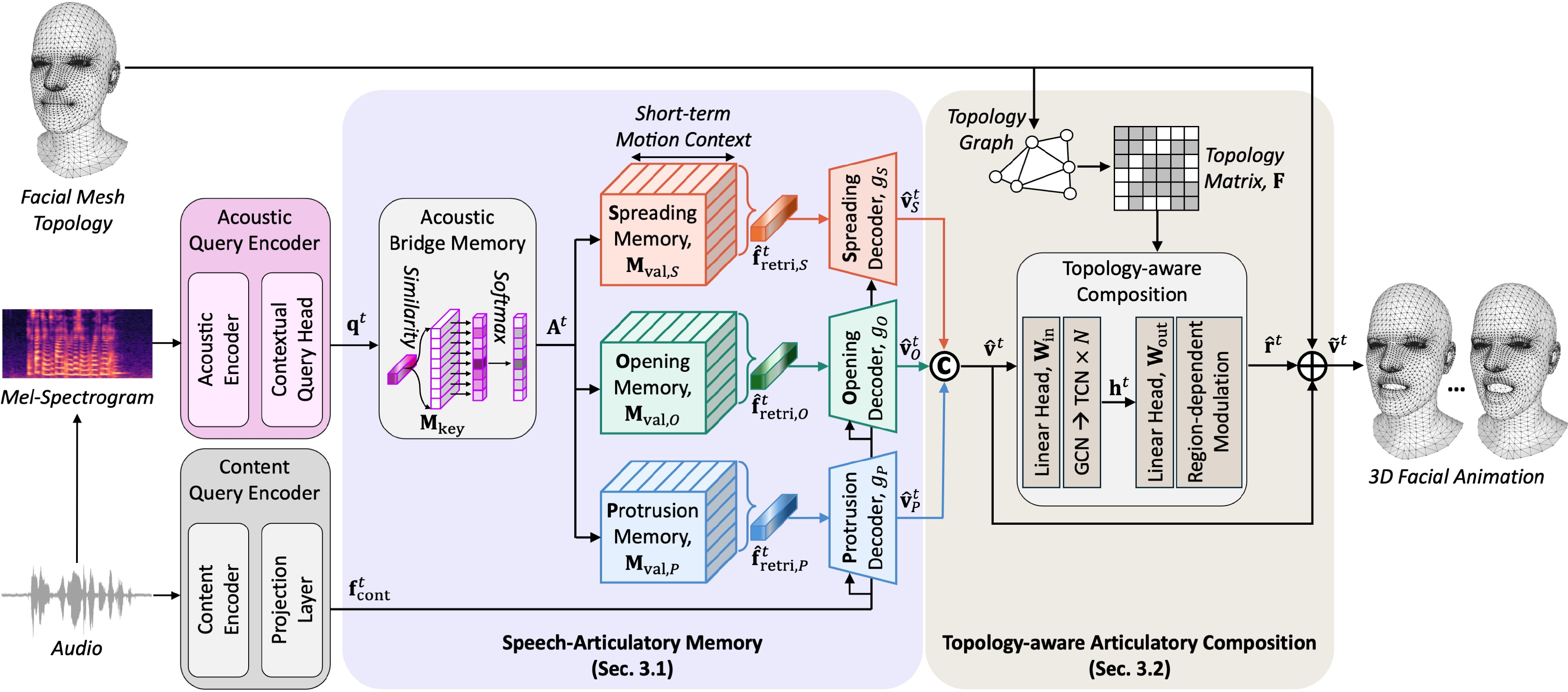}
\end{center}
\vspace{-6pt}
\caption{
    Overall architecture of our articulation-aware framework.
    SAM retrieves directional articulatory motion features (Spreading--Rounding, Opening--Closing, Protrusion--Retraction) from speech.
    TAC composes them into surface-consistent 3D facial motion under the mesh topology.
}
\label{fig:method}
\vspace{-10pt}
\end{figure*}

\section{Proposed Method}
\label{bc1}
Fig.~\ref{fig:method} shows an overview of our articulation-aware speech-driven 3D facial animation framework.
Given a facial template mesh with $V$ vertices, we represent visible articulation through three directional articulatory motions grounded in speech production: horizontal spreading ($S$), vertical opening ($O$), and depth-wise protrusion ($P$)~\cite{lofqvist2005lip, lucero2008analysis, wang2013articulatory, chartier2018encoding}.
We model facial motion $\mathbf{v}^t\in \mathbb{R}^{V \times 3}$ at time $t$ through these three directions ($S$, $O$, $P$), corresponding to the $x$-, $y$-, and $z$-components.
To generate physically grounded visible articulation, the proposed method consists of two main parts, which are Speech--Articulatory Memory (SAM) and Topology-aware Articulatory Composition (TAC).
First, SAM retrieves directional articulatory motion features $\mathbf{f}_{\mathrm{retri},\delta}^{t-\Omega:t+\Omega}$ from value memories $\mathbf{M}_{\mathrm{val},\delta}$ using an acoustic query $\mathbf{q}^t$ and an acoustic bridge memory $\mathbf{M}_{\mathrm{key}}$. 
Then these features are refined via cross-attention with content features $\mathbf{f}_{\mathrm{cont}}^t$ to predict directional articulatory motion $\hat{\mathbf{v}}_{\delta}^t$ for each direction $\delta$.
Second, TAC composes the refined directional articulatory motions under the facial mesh topology to produce surface-consistent 3D facial motion $\tilde{\mathbf{v}}^t$.

\subsection{Speech--Articulatory Memory (SAM)}
\paragraph{\textbf{Acoustic-to-Articulatory Motion Memory Addressing.}} 
Acoustic bridge memory $\mathbf{M}_{\mathrm{key}}\in\mathbb{R}^{N\times D_{\mathrm{SAM}}}$ ($N$ learnable slots of dimension $D_{\mathrm{SAM}}$) is designed to associate speech acoustics with articulatory motion patterns.
We first encode the mel-spectrogram into an acoustic query $\mathbf{q}^t \in \mathbb{R}^{1\times D_{\mathrm{SAM}}}$ using an acoustic encoder with a contextual query head that aggregates a temporal window of size $2\Omega+1$ around time $t$ to capture context-dependent phonetic cues~\cite{craig2008linear, kim25r_interspeech}.
We then compute the addressing vector $\mathbf{A}^t \in \mathbb{R}^{1\times N}$ via cosine similarity followed by scaled softmax with $\tau_{\mathrm{SAM}}{=}16.0$.
\begin{equation}
    \mathbf{A}_{j}^t = \frac{\exp(s_{j}^t\cdot\tau_{\mathrm{SAM}})}{\sum_{k=1}^{N} \exp(s_{k}^t\cdot\tau_{\mathrm{SAM}})}, 
    \quad \text{where } s_{j}^t = \frac{\mathbf{q}^{t} \mathbf{M}_{\mathrm{key},j}^{\mathsf{T}}}{\lVert \mathbf{q}^{t}\rVert_2 \cdot \lVert \mathbf{M}_{\mathrm{key},j}\rVert_2}, 
    \quad j=1,\dots,N.
\end{equation}

\paragraph{\textbf{Directional Articulatory Motion Retrieval.}}  
Given the addressing vector $\mathbf{A}^t$, we retrieve directional motion features from three learnable value memories $\mathbf{M}_{\mathrm{val},\delta}\in \mathbb{R}^{N\times (2\Omega+1) \times D_{\mathrm{SAM}}}$ corresponding to \textit{Spreading} ($S$), \textit{Opening} ($O$), and \textit{Protrusion} ($P$).
Each memory slot stores a directional motion feature sequence of length $2\Omega+1$, encoding short-term temporal continuity.
For each articulatory direction $\delta$, a motion feature sequence $\mathbf{f}_{\mathrm{retri},\delta}^{t-\Omega:t+\Omega} \in \mathbb{R}^{(2\Omega+1) \times D_{\mathrm{SAM}}}$ is retrieved from the corresponding value memory using the addressing vector $\mathbf{A}^t$ via weighted summation, which can be written as
\begin{equation}
    \mathbf{f}_{\mathrm{retri},\delta}^{t-\Omega:t+\Omega} = \sum_{j=1}^{N} \mathbf{A}_{j}^t \cdot \mathbf{M}_{\mathrm{val},\delta,j},
\end{equation}
where $\Omega$ is the half-window defining the temporal span of local phonetic context. 
The retrieved sequence $\mathbf{f}_{\mathrm{retri},\delta}^{t-\Omega:t+\Omega}$ represents short-term articulatory motion features.
\paragraph{\textbf{Content-guided Motion Refinement.}}
To disambiguate phoneme-dependent articulatory patterns, we refine the retrieved motion features by conditioning on content representations from a self-supervised speech model.
Following~\cite{FaceXHuBERT_ICMI23}, we extract a content feature $\mathbf{f}_{\mathrm{cont}}^t \in \mathbb{R}^{1\times D_{\mathrm{SAM}}}$ for $t = 1, \cdots, T$ using a content encoder~\cite{HuBERT2021}.
The extracted features are resampled to match the dataset frame rate and then projected via a linear layer.
Using the content feature sequence $\mathbf{f}_{\mathrm{cont}}^{1:T}$ as the query and $\hat{\mathbf{f}}_{\mathrm{retri},\delta}^{1:T}$, summarized from retrieved sequences, as key/value, a per-direction Transformer decoder $g_{\delta}(\cdot)$ performs temporal cross-attention to produce the directional articulatory motion sequence $\hat{\mathbf{v}}_\delta^{1:T}$.
\begin{equation}
    \hat{\mathbf{v}}_\delta^{1:T} = g_{\delta}(\mathbf{f}_{\mathrm{cont}}^{1:T}, \hat{\mathbf{f}}_{\mathrm{retri},\delta}^{1:T}) = \mathrm{softmax}\!\left( \frac{\mathbf{Q}_\delta \mathbf{K}_\delta^{\mathsf{T}}}{\sqrt{D_{\mathrm{SAM}}}} \right) \mathbf{V}_\delta \, \mathbf{W}_{\mathrm{out},\delta},
\end{equation}

\noindent where $\mathbf{Q}_\delta = (\mathbf{f}_{\mathrm{cont}}^{1:T} + \mathbf{P})\mathbf{W}_{Q,\delta}$ with periodic positional encoding $\mathbf{P}$~\cite{faceformer2022} of period $\rho$ to encode temporal periodicity (see Appendix~\ref{supp:ppe} for details), $\mathbf{K}_\delta = \hat{\mathbf{f}}_{\mathrm{retri},\delta}^{1:T} \mathbf{W}_{K,\delta}$, and $\mathbf{V}_\delta = \hat{\mathbf{f}}_{\mathrm{retri},\delta}^{1:T} \mathbf{W}_{V,\delta}$.
Here, $\mathbf{W}_{Q,\delta}, \mathbf{W}_{K,\delta}, \mathbf{W}_{V,\delta} \in \mathbb{R}^{D_{\mathrm{SAM}} \times D_{\mathrm{SAM}}}$ and $\mathbf{W}_{\mathrm{out},\delta} \in \mathbb{R}^{D_{\mathrm{SAM}} \times V}$ are learnable projection matrices.
$\hat{\mathbf{f}}_{\mathrm{retri},\delta}^t = \sum_{\omega=-\Omega}^{+\Omega} w_{\delta}(\omega) \cdot \mathbf{f}_{\mathrm{retri},\delta}^{t+\omega}$ summarizes the retrieved sequence via learnable weights $w_{\delta}(\omega)$.


\paragraph{\textbf{Training Objectives.}}
We train the Speech--Articulatory Memory (SAM) to retrieve and synthesize directional articulatory motion $\hat{\mathbf{v}}_\delta^t$ by aligning with the corresponding ground-truth motion $\mathbf{v}_\delta^t$. 
To this end, we apply a directional reconstruction loss $\mathcal{L}_{\mathrm{rec},\delta}$ to enforce accurate prediction of the ground-truth motion $\mathbf{v}_\delta^t$ in the vertex domain.
In addition, a directional velocity loss $\mathcal{L}_{\mathrm{vel},\delta}$ on temporal differences $\Delta\mathbf{v}_\delta^t$ is also employed to encourage the predicted motion to follow the correct temporal evolution.
These complementary supervisions jointly enforce accurate spatial reconstruction and temporally consistent motion dynamics for each articulatory direction $\delta$.
\begin{equation}
\label{main:vertex_domain_supervisions}
    \mathcal{L}_{\mathrm{rec},\delta} = \frac{1}{T}\sum_{t=1}^{T}\lVert \mathbf{v}_\delta^t-\hat{\mathbf{v}}_\delta^t \rVert_2^{2}, \quad
    \mathcal{L}_{\mathrm{vel},\delta} = \frac{1}{T-1}\sum_{t=1}^{T-1}\lVert \Delta\mathbf{v}_\delta^t-\Delta\hat{\mathbf{v}}_\delta^t \rVert_2^{2},
\end{equation}
where $\Delta\mathbf{v}_\delta^t = \mathbf{v}_\delta^{t+1} - \mathbf{v}_\delta^t$. These supervisions mitigate shortcut compensation across directions and encourage each directional motion to follow its intended spatial and temporal behavior.

We further introduce a value memory store loss $\mathcal{L}_{\mathrm{store},\delta}$ to guide the value memory to store temporally coherent articulatory motion patterns.
For this purpose, we align the retrieved motion feature sequence $\mathbf{f}_{\mathrm{retri},\delta}^{t-\Omega:t+\Omega}$ with the reference motion feature sequence $\mathbf{z}_{\delta}^{t-\Omega:t+\Omega}$ defined over a $(2\Omega+1)$-frame temporal window.
For each articulatory direction $\delta$, the reference motion feature $\mathbf{z}^t_{\delta} \in \mathbb{R}^{1\times D_{\mathrm{SAM}}}$ is extracted from ground-truth motion $\mathbf{v}_{\delta}^t$ using a pre-trained per-direction autoencoder, which is trained to encode $\mathbf{v}_{\delta}^t$ into $\mathbf{z}^t_{\delta}$ and reconstruct $\mathbf{v}_{\delta}^t$ from $\mathbf{z}^t_{\delta}$ (see Appendix~\ref{supp:archi_detail} for details).
At each time step $t$, we align the reference window $\mathbf{z}_{\delta}^{t-\Omega:t+\Omega}$ with the  $\mathbf{f}_{\mathrm{retri},\delta}^{t-\Omega:t+\Omega}$ via cosine distance:
\begin{equation}
\label{main:store_loss}
    \mathcal{L}_{\mathrm{store},\delta} = \frac{1}{T\cdot(2\Omega+1)}\sum_{t=1}^{T} \sum_{\omega=-\Omega}^{+\Omega} \left( 1 - \frac{\mathbf{f}_{\mathrm{retri},\delta}^{t+\omega} \cdot \mathbf{z}_{\delta}^{t+\omega}}{\lVert \mathbf{f}_{\mathrm{retri},\delta}^{t+\omega} \rVert_2 \lVert \mathbf{z}_{\delta}^{t+\omega} \rVert_2} \right).
\end{equation}
The window-level alignment encourages the value memory to capture short-term temporal consistency in articulatory motion, reducing one-to-many ambiguity in acoustic-to-motion mapping.

The total loss for SAM is $\mathcal{L}_{\mathrm{SAM}} = \sum_{\delta \in \{S,O,P\}}\left(\mathcal{L}_{\mathrm{rec},\delta} + \mathcal{L}_{\mathrm{vel},\delta} + \lambda_{1}\mathcal{L}_{\mathrm{store},\delta}\right)$, with $\lambda_{1} = 10^{-6}$. 

\subsection{Topology-aware Articulatory Composition (TAC)}
\label{sec:tac}
In TAC, the predicted articulatory motions $\hat{\mathbf{v}}_{\delta}^t$ are composed into surface-consistent 3D facial motion by considering their spatio-temporal coordination over the facial surface.
To this end, we introduce a topology-aware composition that captures spatial coupling across vertices and temporal consistency.

\paragraph{\textbf{Spatio-Temporal Composition via Mesh Topology.}}
For spatial coordination, we leverage the facial topology adjacency matrix $\mathbf{F}\in\mathbb{R}^{V\times V}$ derived from the facial template mesh.
$\mathbf{F}_{ij}=1$ if vertices $i$ and $j$ are connected by an edge, and $\mathbf{F}_{ij}=0$ otherwise.
Let $\mathbf{h}^{t} \in\mathbb{R}^{V \times D_{\mathrm{TAC}}}$ denote the topology-aware motion feature. 
We first stack the refined articulatory motions $\hat{\mathbf{v}}_{\delta}^t$ for each direction $\delta$ as $\hat{\mathbf{v}}^{t}=[\hat{\mathbf{v}}_S^t,~\hat{\mathbf{v}}_O^t,~\hat{\mathbf{v}}_P^t]\in\mathbb{R}^{V\times 3}$ and project it into a latent space $\hat{\mathbf{h}}^{t} = \hat{\mathbf{v}}^{t}\mathbf{W}_{\mathrm{in}} \in \mathbb{R}^{V\times D_{\mathrm{TAC}}}$, where $\mathbf{W}_{\mathrm{in}} \in \mathbb{R}^{3\times D_{\mathrm{TAC}}}$.
Then, $\mathbf{h}^{t}$ can be obtained by propagating motion information across neighboring vertices with a graph convolutional layer~\cite{kipf2017semi}.
\begin{equation}
    \mathbf{h}^{t} ~=~ \sigma\left(\hat{\mathbf{F}}\cdot \hat{\mathbf{h}}^{t}\cdot \mathbf{W} \right),
\end{equation}
where $\hat{\mathbf{F}}=\tilde{\mathbf{D}}^{-1/2}(\mathbf{F}+\mathbf{I})\tilde{\mathbf{D}}^{-1/2}$ is the symmetrically normalized adjacency matrix following~\cite{kipf2017semi} and $\sigma(\cdot)$ is a non-linear activation. 
$\tilde{\mathbf{D}}$ is the degree matrix of $\mathbf{F}+\mathbf{I}$, and $\mathbf{W}\in\mathbb{R}^{D_{\mathrm{TAC}}\times D_{\mathrm{TAC}}}$ is a learnable transformation.
This operation injects mesh-topology information into $\mathbf{h}^{t}$, allowing each vertex representation to reflect coordinated motions over its topological neighborhood.

To capture temporal dynamics, we model the topology-aware motion features as vertex-wise time-varying sequences and apply a temporal convolutional network (TCN) along the time axis.
For each vertex $i$, the topology-aware motion feature sequence $\mathbf{h}_{i}^{1:T}\in\mathbb{R}^{T\times D_{\mathrm{TAC}}}$ is defined as
\begin{equation}
    \mathbf{h}_{i}^{1:T} ~\leftarrow~ \mathrm{TCN}\left(\mathbf{h}_{i}^{1:T}\right), \quad i=1,\dots,V.
\end{equation}
This temporal composition allows each vertex to capture short-term temporal dependencies while preserving spatial coordination from the mesh topology, resulting in consistent facial motion.
Finally, we project the topology-aware motion feature sequences $\mathbf{h}^{1:T}$ to the vertex domain via a linear layer $\mathbf{W}_{\mathrm{out}}\in\mathbb{R}^{D_{\mathrm{TAC}}\times 3}$ yielding topology-aware motion residual $\mathbf{r}^{t}=\mathbf{h}^{t}\mathbf{W}_{\mathrm{out}}\in\mathbb{R}^{V\times 3}$.

\paragraph{\textbf{Region-adaptive Motion Modulation.}}
Facial regions exhibit different deformation characteristics (e.g., highly deformable lips/jaw vs.\ rigid cheeks/forehead).
To reflect such region-dependent deformability, we introduce a learnable per-vertex modulation gate 
$\mathbf{m}\in\mathbb{R}^{V\times1}$, where $\mathbf{m}=V\cdot\mathrm{softmax} (\tau_{\mathrm{TAC}}\cdot\boldsymbol{\alpha})$.
$\boldsymbol{\alpha}\in\mathbb{R}^{V\times1}$ is a learnable vector and $\tau_{\mathrm{TAC}}$ is a temperature constant set to 16.0.
We broadcast $\mathbf{m}$ along the coordinate dimension to obtain a region-modulated motion residual.
\begin{equation}
    \hat{\mathbf{r}}^{t} = (\mathbf{m}\cdot \mathbbm{1})\odot \mathbf{r}^{t},
\end{equation}
where $\mathbbm{1}\in\mathbb{R}^{1\times3}$ is an all-ones vector and $\odot$ denotes element-wise multiplication.

By modulating $\mathbf{r}^{t}$ in a region-adaptive manner, the gate $\mathbf{m}$ allows different facial regions to absorb the topology-aware motion residual with different sensitivities.
Finally, the surface-consistent 3D facial motion is obtained by $\tilde{\mathbf{v}}^{t} = \hat{\mathbf{v}}^{t} + \hat{\mathbf{r}}^{t}$.
\paragraph{\textbf{Training Objectives.}}
We train the TAC to learn a topology-aware refinement that composes the directional articulatory motions into surface-consistent 3D facial motion.
$\mathbf{v}^{t}\in\mathbb{R}^{V\times 3}$ denotes the ground-truth facial motion field at $t$. 
Let $\mathbf{L} = \mathbf{I} - \hat{\mathbf{F}}$ denote the normalized graph Laplacian derived from the mesh topology.
The reconstruction, velocity, and graph Laplacian losses can be defined as
\begin{equation}
    \begin{aligned}
        \mathcal{L}_{\mathrm{rec}}
        = \tfrac{1}{T}\sum_{t=1}^{T}\lVert \mathbf{v}^{t}-\tilde{\mathbf{v}}^{t}\rVert_{2}^{2},\quad
        \mathcal{L}_{\mathrm{vel}} = \tfrac{1}{T-1}\sum_{t=1}^{T-1}
        \lVert \Delta\mathbf{v}^{t} - \Delta\tilde{\mathbf{v}}^{t} \rVert_{2}^{2}, \quad
        \mathcal{L}_{\mathrm{lap}} = \tfrac{1}{TV}\sum_{t=1}^{T}\lVert \mathbf{L}\mathbf{v}^{t} - \mathbf{L}\tilde{\mathbf{v}}^{t} \rVert_{2}^{2},
    \end{aligned}
\end{equation}
While $\mathcal{L}_{\mathrm{rec}}$ and $\mathcal{L}_{\mathrm{vel}}$ enforce per-vertex spatial and temporal accuracy, $\mathcal{L}_{\mathrm{lap}}$ measures each vertex's deviation from the topology-weighted average of its neighbors, enforcing local surface consistency.
The total TAC loss is $\mathcal{L}_{\mathrm{TAC}} = \mathcal{L}_{\mathrm{rec}} + \mathcal{L}_{\mathrm{vel}} + \lambda_{2}\mathcal{L}_{\mathrm{lap}}$, with $\lambda_{2} = 10^{3}$.



\section{Experiments}
\subsection{Datasets}
\label{bc2}
We evaluate on two widely used datasets, VOCASET~\cite{VOCA2019} and TFHP~\cite{diffposetalk2024}.
VOCASET consists of 480 sequences from 12 speakers at 60 fps in FLAME topology~\cite{FLAME2017} (5,023 vertices), with an 8/2/2 speaker split for train/val/test.
Following standard protocols~\cite{faceformer2022, codetalker2023}, we use its 255 unique sentences for evaluation.
TFHP contains 1,052 videos (26.5 hours, including 348 from HDTF~\cite{zhang2021flow}) from 588 speakers, covering lectures, interviews, and news programs.
All videos are converted to 25 fps, yielding $\sim$2.38M frames with no-head-pose FLAME parameters~\cite{diffposetalk2024} (decoded into vertex motion under the FLAME topology). 
The data is split 460/64/64 by speakers for train/val/test.

\subsection{Implementation Details}
\label{bc3}
All experiments are conducted on a single NVIDIA RTX 4090 GPU. 
We adopt a two-stage training strategy with the Adam optimizer (learning rate $10^{-4}$, batch size 1).
In Stage~1, we train SAM for 250 epochs with the Wav2Lip~\cite{wav2lip} mel-spectrogram encoder and pre-trained HuBERT~\cite{HuBERT2021} as the content encoder.
In Stage~2, we freeze SAM and train TAC for 150 epochs.
We use $N{=}32$ memory slots, $\Omega{=}7$, $D_{\mathrm{SAM}}{=}128$, and $D_{\mathrm{TAC}}{=}64$.
$\hat{\mathbf{F}}$ is pre-computed from the dataset-specific mesh topology and kept fixed during training.
We apply zero padding for temporal boundary handling.
Architecture details are provided in Appendix~\ref{supp:archi_detail}.

\subsection{Evaluation Metrics}
\label{bc4}
We use four standard metrics: \textbf{FVE} (mean Euclidean distance over all vertices)~\cite{faceformer2022}, \textbf{LVE} (per-frame max $\ell_2$ error on lip vertices)~\cite{faceformer2022,codetalker2023}, \textbf{FDD} (upper-face dynamics deviation)~\cite{codetalker2023}, and \textbf{LDTW} (DTW~\cite{dtw07}-based lip alignment)~\cite{imitator2023}.

In addition, we introduce visible articulatory distance metrics computed from lip regions shown in Fig.~\ref{fig_1_thumb}(a).
Let $\mathbf{p}_{\mathrm{UL}}(t),\ \mathbf{p}_{\mathrm{LL}}(t),\ 
\mathbf{p}_{\mathrm{LMC}}(t),$ and $\mathbf{p}_{\mathrm{RMC}}(t) \in \mathbb{R}^3$ denote the region-mean 3D positions of UL, LL, LMC, and RMC at frame $t$, respectively.
We define three articulatory measures as
\begin{equation}
\begin{gathered}
\text{ICW}(t) = |\mathbf{p}_{\mathrm{LMC},x}(t) - \mathbf{p}_{\mathrm{RMC},x}(t)|, \quad \\
\text{ILD}(t) = |\mathbf{p}_{\mathrm{UL},y}(t) - \mathbf{p}_{\mathrm{LL},y}(t)|, \\
\text{LP}(t)  = (\mathbf{p}_{\mathrm{UL},z}(t) - \mathbf{p}_{\mathrm{UL},z}^{\mathrm{neu}}) + (\mathbf{p}_{\mathrm{LL},z}(t) - \mathbf{p}_{\mathrm{LL},z}^{\mathrm{neu}}),
\end{gathered}
\end{equation}
where ICW measures the inter-commissural width (lip spreading) and ILD measures the inter-labial distance (lip opening).
LP captures the lip protrusion relative to the neutral template positions $\mathbf{p}_{\mathrm{UL},z}^{\mathrm{neu}}$ and $\mathbf{p}_{\mathrm{LL},z}^{\mathrm{neu}}$.
We evaluate articulation accuracy using \textbf{Visible Articulatory Distance Error} (RMSE on distance) and \textbf{Visible Articulatory Velocity Error} (RMSE on temporal derivatives).

\subsection{Quantitative Results}
\begin{wraptable}{r}{0.55\textwidth}
  \vspace{-55pt}
  \centering
  \small
  \setlength{\tabcolsep}{2pt}
  \renewcommand{\arraystretch}{1.05}
  \newcommand{\mw}{0.06\textwidth}
    \caption{Quantitative results on VOCASET and TFHP.}
    \vspace{-5pt}
  \begin{tabular}{ll w{c}{\mw} w{c}{\mw} w{c}{\mw} w{c}{\mw}}
    \toprule
    \multirow{2}{*}{Dataset} 
    & \multirow{2}{*}{Methods} 
    & \multirow{2}{*}{FVE$\downarrow$} 
    & \multirow{2}{*}{LVE$\downarrow$} 
    & \multirow{2}{*}{FDD$\downarrow$}
    & \multirow{2}{*}{LDTW$\downarrow$} \\
    & & & & & \\
    \midrule
    
    \multirow{9}{*}{VOCASET}
    & FaceFormer                         & 0.878 & 0.346 & 0.143 & 0.186 \\
    & CodeTalker                         & 0.913 & 0.397 & 0.160 & 0.204 \\
    & SelfTalk                           & 0.844 & 0.305 & 0.148 & 0.174 \\
    & Imitator                           & 0.898 & 0.384 & 0.179 & 0.199 \\
    & ScanTalk                           & 0.878 & 0.310 & 0.152 & 0.168 \\
    & UniTalker                          & 0.820 & 0.282 & \underline{0.101}    & 0.188 \\
    & MemoryTalker                       & {0.810} & \underline{0.250} & 0.127  & \underline{0.151} \\
    & StreamingTalker                    & \underline{0.782} & {0.272} & 0.191  & 0.169 \\
    \rowcolor{gray!20}
    \cellcolor{white} & \textbf{Ours}    & \textbf{0.771} & \textbf{0.235} & \textbf{0.097} & \textbf{0.132} \\
    \midrule
    
    \multirow{4}{*}{TFHP}
    & Mimic                             & \underline{0.106} & \underline{0.116} & 0.619 & \underline{0.329} \\
    & DiffPoseTalk                      & {0.176} & {0.159} & \textbf{0.321} & {0.382} \\
    & ARTalk                            & 0.206 & 0.164 & 0.581 & 0.385 \\
    \rowcolor{gray!20}
    \cellcolor{white} & \textbf{Ours}   & \textbf{0.100} & \textbf{0.108} & \underline{0.578} & \textbf{0.318} \\
    \bottomrule
  \end{tabular}
  \label{table1:quantitative_evaluation}
  \vspace{-30pt} 
\end{wraptable}
\subsubsection{Overall Comparisons}
As shown in Table~\ref{table1:quantitative_evaluation}, our method achieves the lowest FVE, LVE, and LDTW on both datasets.
The LVE and LDTW gains indicate that articulatory modeling improves lip-region reconstruction, while the lowest FVE demonstrates that overall reconstruction quality is preserved.
Results on the TFHP demonstrate consistent lip-region improvements without degrading overall reconstruction quality across various speaking styles.
But our method does not explicitly consider upper-face dynamics for its impact on FDD (see Sec.~\ref{sec:limitations}).

\subsubsection{Articulatory Motion Comparisons}
\begin{wraptable}{r}{0.55\textwidth}
  \vspace{-12pt}
  \centering
  \small
  \setlength{\tabcolsep}{2pt}
  \renewcommand{\arraystretch}{1.05}
  \caption{Visible articulatory distance and velocity errors on VOCASET.}
  \vspace{-5pt}
  \begin{tabular}{lccc|ccc}
    \toprule
    \multirow{2}{*}{Methods}
    & \multicolumn{3}{c|}{Distance Error}
    & \multicolumn{3}{c}{Velocity Error} \\
    & ICW$\downarrow$ & ILD$\downarrow$ & LP$\downarrow$
    & ICW$\downarrow$ & ILD$\downarrow$ & LP$\downarrow$ \\
    \midrule
    FaceFormer       & 1.764 & 4.558 & 2.429 & 0.581 & 1.850 & 1.230 \\
    CodeTalker       & 1.725 & 4.777 & 2.808 & 0.552 & 2.009 & 1.343 \\
    SelfTalk         & 1.680 & 4.031 & 2.470 & 0.544 & 1.958 & 1.275 \\
    Imitator         & 1.731 & 4.851 & 2.526 & 0.571 & 1.873 & {1.205} \\
    ScanTalk         & 1.737 & 4.034 & 2.485 & 0.558 & 1.854 & 1.285 \\
    UniTalker        & \underline{1.585} & 3.887 & 2.532 & {0.541} & \underline{1.627} & 1.211 \\
    MemoryTalker     & {1.626} & \underline{3.679} & \underline{2.347} & 0.560 & 1.813 & 1.224 \\
    StreamingTalker  & 1.658 & 3.955 & 2.478 & \textbf{0.507} & 1.645 & \underline{1.187} \\
    \rowcolor{gray!20}
    \textbf{Ours}    & \textbf{1.568} & \textbf{3.407} & \textbf{2.346} & \underline{0.521} & \textbf{1.613} & \textbf{1.177} \\
    \bottomrule
  \end{tabular}
  \label{table2:articulator_distance_velocity_errors}
  \vspace{-20pt}
\end{wraptable}
To measure articulatory motion, we use RMSE-based distance and velocity errors on three articulatory measures, which are ICW, ILD, and LP.
As shown in Table~\ref{table2:articulator_distance_velocity_errors}, our method achieves the lowest distance errors on all three measures, indicating that the directional decomposition accurately captures the spatial magnitude of articulatory motion.
For velocity errors, our method performs the best on ILD and LP and the second-best on ICW, suggesting that this spatial accuracy is accompanied by consistent temporal dynamics.
Additional TFHP results are provided in Sec.~\ref{sec:articulatory_error_tfhp}.

\begin{wrapfigure}{r}{0.55\textwidth} 
  \vspace{-3pt} 
  \centering
  \includegraphics[width=1.0\linewidth]{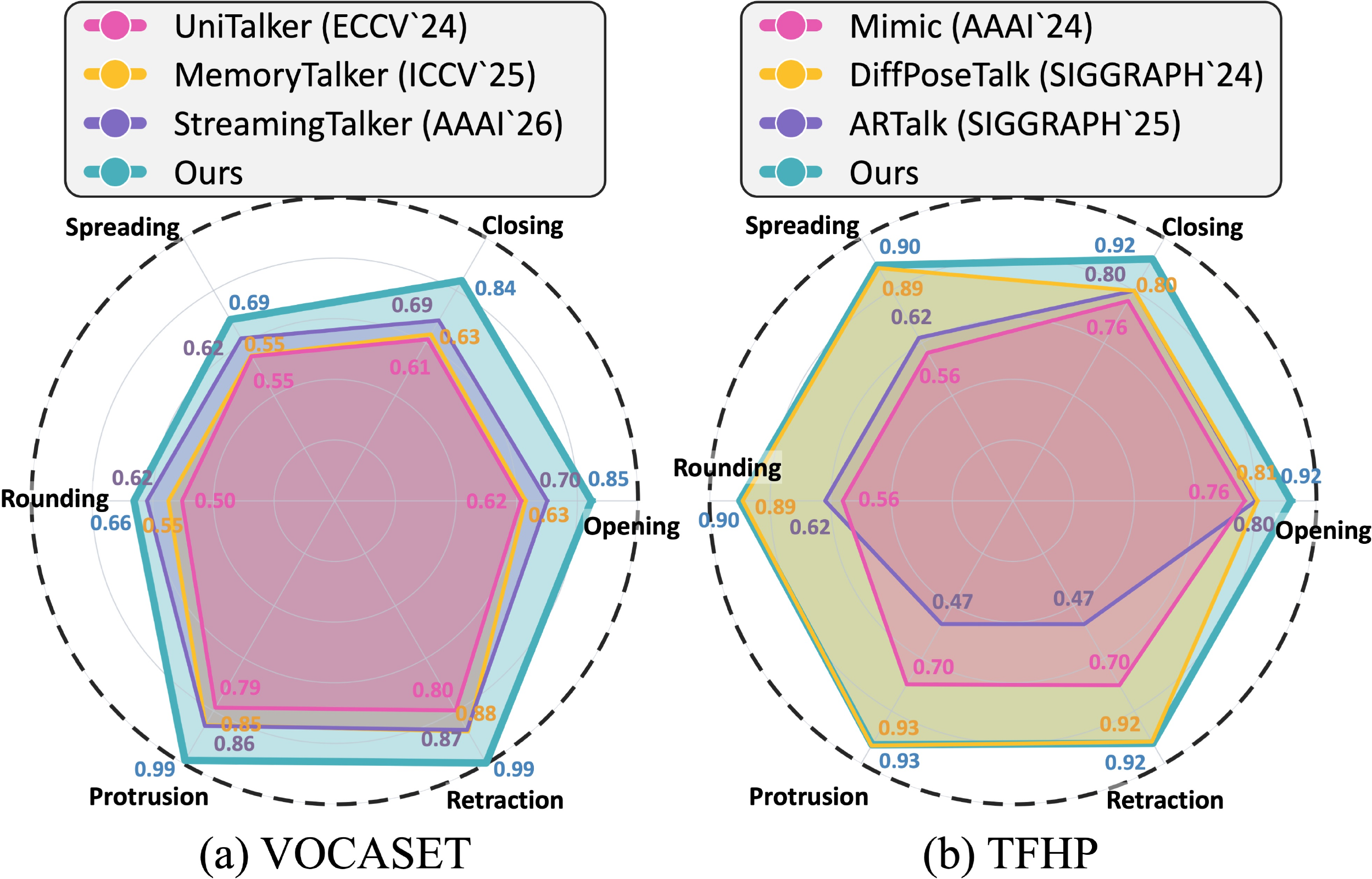}
  \vspace{-15pt} 
  \caption{Radar plot of directional lip-motion tendency strength relative to ground truth.}
  \label{fig3:radar_plot}
  \vspace{-5pt} 
\end{wrapfigure}
Beyond per-frame accuracy, we evaluate whether each method preserves the relative magnitude of directional articulatory motions.
For each direction $\delta$, we derive a 1D kinematic trajectory from the lip anchors (UL, LL, LMC, RMC) and compute its temporal slope.
We decompose the slope into positive and negative components, corresponding to paired motions (e.g., Opening--Closing), and measure the magnitude of each motion as the mean absolute value of its component over time.
We then compare the predicted and ground-truth magnitudes using $\exp(-|\log(r)|)$, where $r$ is their ratio (equal to 1 when matched, decaying symmetrically otherwise).
See Appendix~\ref{supp:tendency_analysis} for details.

Fig.~\ref{fig3:radar_plot} shows the results as radar plots normalized by ground truth. 
Our method remains closest to the ground truth across all directions, indicating more balanced directional articulation.
\begin{figure}[t!]
\centering
  \vspace{-10pt}
\includegraphics[width=0.95\linewidth]{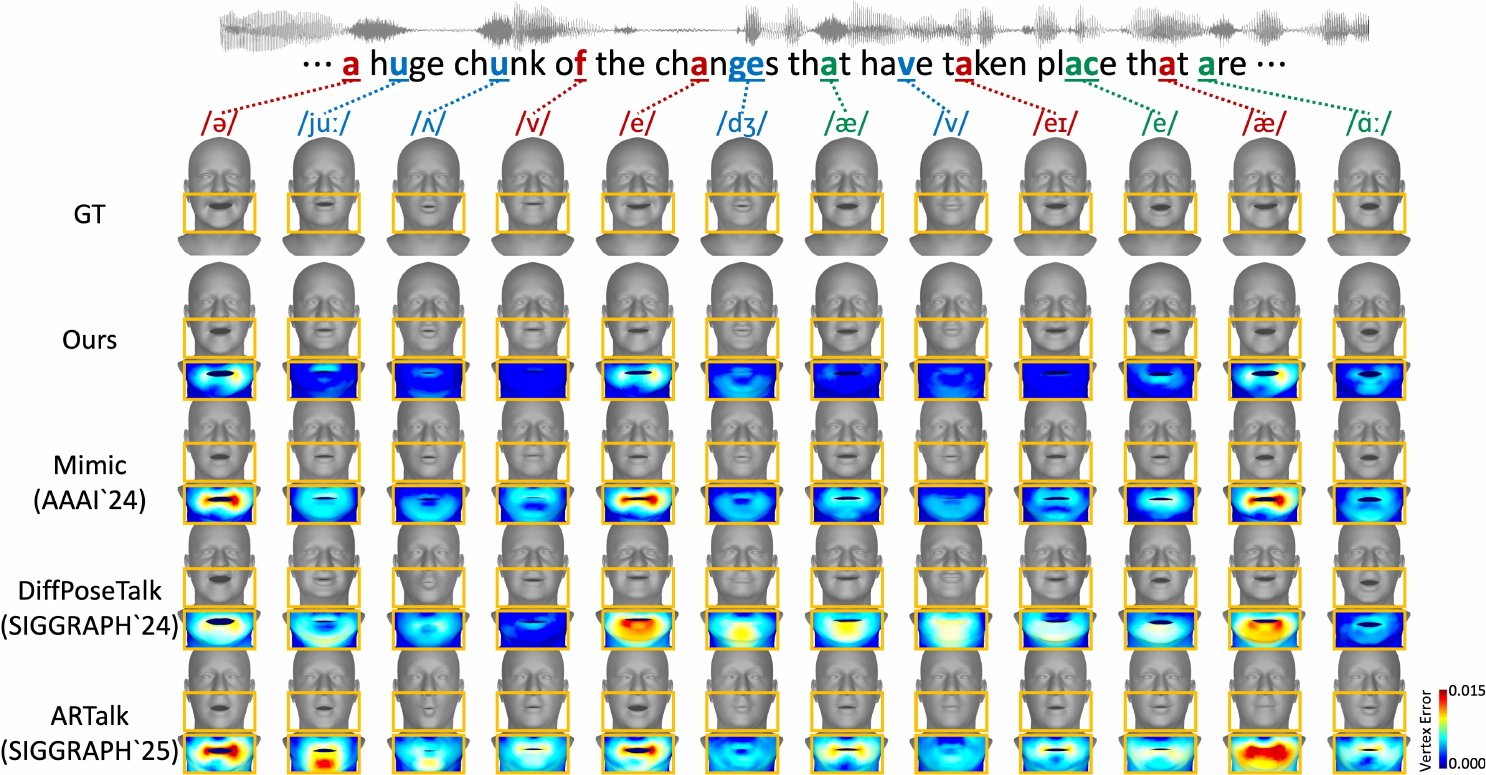}
\vspace{-5pt}
\caption{Sequence-level qualitative results on TFHP. 
For each method, mesh renderings (top) and per-vertex L1 error against the ground truth (bottom, color-coded from 0 (blue) to 0.015 (red)) are shown over time at phoneme-aligned frames.}
\vspace{-10pt}
\label{fig4:qualitative_eval}
\end{figure}

\begin{figure}[t!]
  \centering
  \includegraphics[width=0.95\linewidth]{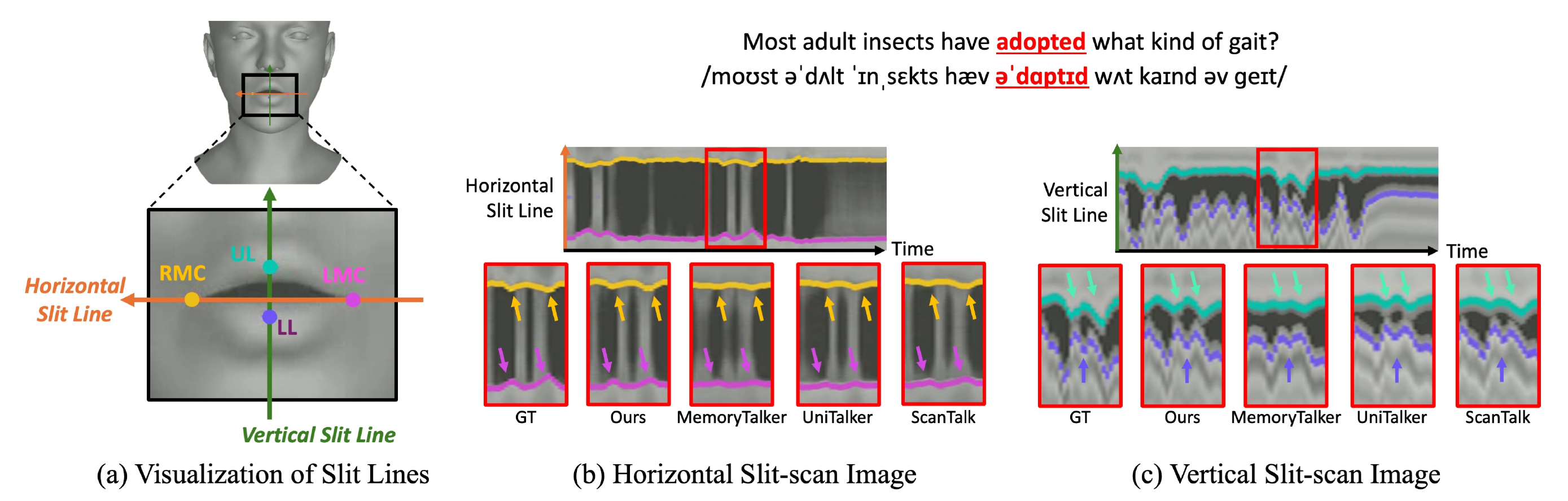}
  \vspace{-5pt}
  \caption{Slit-scan visualizations for sequence-level comparison.
  (a) lip ROI and landmarks (UL, LL, LMC, RMC) defining horizontal (LMC--RMC) and vertical (UL--LL) slit lines.
  (b) Horizontal and (c) Vertical slit-scans over time, reflecting lip spreading (width) and lip opening (height), respectively.}
  \label{fig5:slit_scan}
  \vspace{-10pt}
\end{figure}
\subsection{Qualitative Results}

\subsubsection{Sequence-Level Mesh Comparisons}
Fig.~\ref{fig4:qualitative_eval} shows mesh renderings and per-vertex L1 error maps over an utterance on TFHP.
Our method produces distinct articulatory configurations at salient phonemes, supported by low per-vertex L1 errors across the lips and lower-face region.
This indicates that our articulation-aware modeling reconstructs phoneme-specific motion accurately throughout the utterance.
In the mesh renderings, the upper-face region remains comparable to competing methods, indicating that our model captures articulatory motion without affecting non-articulatory regions.

\subsubsection{Sequence-Level Visible Articulatory Dynamics}
Beyond sampling articulation at discrete time points, we further visualize the entire utterance trajectory using slit-scan visualizations ($x$–$t$ and $y$–$t$ slices) in Fig.~\ref{fig5:slit_scan}.
Slit-scan enables direct comparison of these trajectories across methods and exposes the sharpness of articulatory transitions through the curve slope.
For each frame, we average a narrow strip (5 pixels) centered at the horizontal slit (LMC--RMC) and the vertical slit (UL--LL) to obtain 1D profiles, which are stacked over time to form $x$–$t$ and $y$–$t$ maps capturing lip spreading and mouth opening, respectively.
We visualize slit-scan trajectories for the utterance segment \textit{adopted}.
In (b), the LMC and RMC trajectories of our method show distinct rounding and spreading patterns with steeper slopes at articulatory transitions, indicating distinct horizontal articulatory motions that match the ground-truth pattern.
In (c), the UL and LL trajectories of our method show distinct opening and closing patterns with steeper slopes during rapid transitions, indicating more distinct vertical articulatory motions than competing methods.

\subsection{Analysis}
\subsubsection{Topology-aware Composition}

\definecolor{darkgreen}{RGB}{0,130,0}
\definecolor{darkred}{RGB}{180,0,0}
\begin{wraptable}{r}{0.40\linewidth}
\vspace{-50pt}
\centering
\small
\setlength{\tabcolsep}{4pt}
\renewcommand{\arraystretch}{1.05}
\caption{Ablation of TAC module.}
\vspace{-5pt}
\begin{tabular}{ccccc}
\toprule
SAM & TAC & $\text{FVE}\!\downarrow$ & $\text{LVE}\!\downarrow$ & LDTW$\!\downarrow$ \\
\midrule
 \textcolor{darkgreen}{\ding{51}} & \textcolor{darkred}{\ding{55}}   & 0.812 & 0.255 & 0.161 \\
 \rowcolor{gray!20}
 \textcolor{darkgreen}{\ding{51}} & \textcolor{darkgreen}{\ding{51}} & \textbf{0.771} & \textbf{0.235} & \textbf{0.132} \\
\bottomrule
\end{tabular}
\label{table:module_ablation_revised}
\vspace{-10pt}
\end{wraptable}
Table~\ref{table:module_ablation_revised} shows the ablation of the TAC module on VOCASET.
SAM predicts per-direction articulatory motions, each physically grounded in human speech, and reaches between UniTalker and MemoryTalker in Table~\ref{table1:quantitative_evaluation}.
We further introduce TAC to compose the per-direction motions through spatial coupling across the facial surface, improving all metrics and yielding surface-consistent 3D facial animation.
A comparison with a topology-independent composition baseline is provided in Appendix~\ref{supp:tac_ablation}.

\subsubsection{Directional Articulatory Modeling}
\begin{wrapfigure}{r}{0.50\linewidth}
\vspace{-40pt}
\centering
\includegraphics[width=\linewidth]{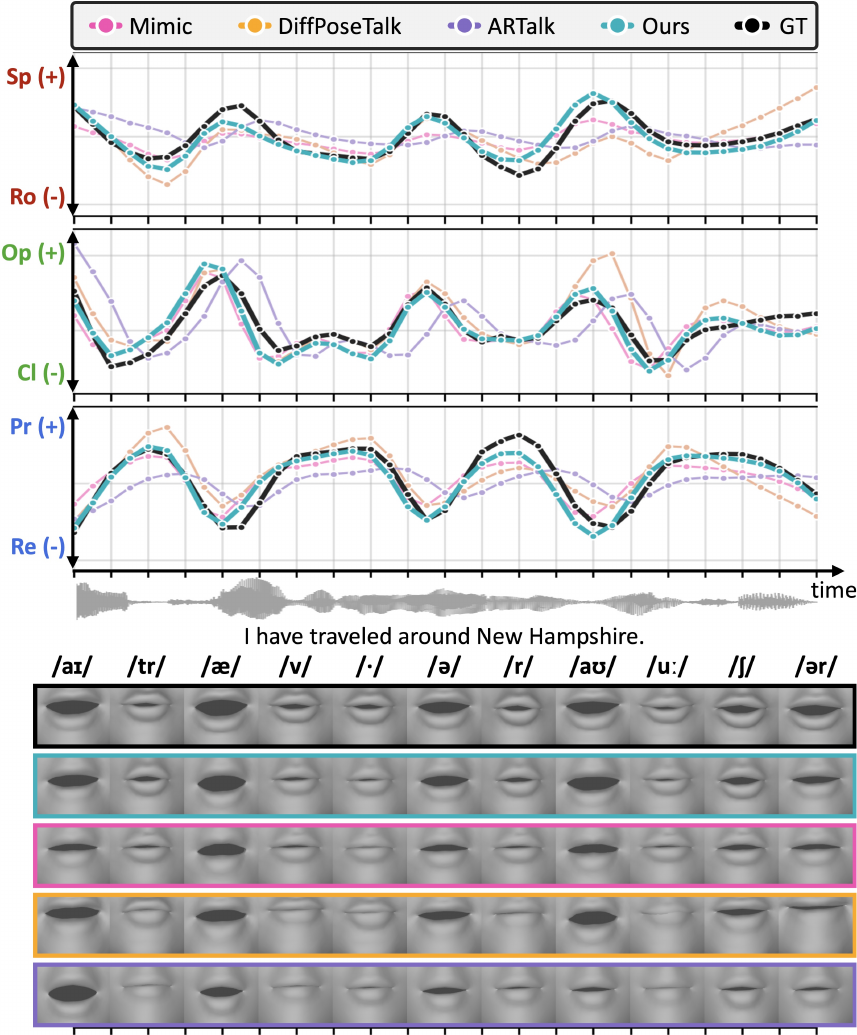}
\vspace{-15pt}
\captionsetup{width=1.15\linewidth}  
\caption{Analysis of Directional articulatory trajectories.}
\vspace{-30pt}
\label{fig6:trajectory_analysis}
\end{wrapfigure}
Fig.~\ref{fig6:trajectory_analysis} visualizes directional articulatory motion trajectories along three directions: Spreading--Rounding, Opening--Closing, and Protrusion--Retraction.
Each trajectory is computed from lip-surface anchors (UL, LL, LMC, RMC), as detailed in Appendix Eq.~(\ref{supp:trajectories}), and aligned with the speech waveform and per-phoneme mesh renderings.
Baseline trajectories deviate from the ground truth in direction-specific ways.
Some show flat Opening--Closing trajectories at opening-dominant phonemes (e.g., /\textipa{aI}/, /\textipa{\ae}/, /\textipa{aU}/), others attenuated Protrusion--Retraction trajectories at protrusion-dominant phonemes (e.g., /tr/, /v/), and others noisy Spreading--Rounding trajectories at high-magnitude segments (e.g., /\textipa{aU}/).
In contrast, our method closely follows the ground truth across all three directions, with consistent reconstruction quality in the mesh. (See Appendix~\ref{ssec:Articulatory_Motion_Trajectory} for more.)

\subsection{User Study}
\label{main:userstudy}
\begin{wraptable}{r}{0.4\textwidth}
  \vspace{-20pt}
  \centering
  \small
  \setlength{\tabcolsep}{2pt}
  \renewcommand{\arraystretch}{1.05}
  \caption{User study: pairwise preference rates (\%) on VOCASET.}
  \vspace{-5pt}
  \begin{tabular}{lcc}
    \toprule
    \multirow{2}{*}{Competitors} & Lip Sync & Realism \\
    & (\% Ours) & (\% Ours) \\
    \midrule
    vs. FaceFormer    & 71.3 & 65.7 \\
    vs. CodeTalker    & 72.2 & 75.0 \\
    vs. Imitator      & 77.8 & 73.1 \\
    vs. ScanTalk      & 73.1 & 79.6 \\
    vs. UniTalker     & 65.7 & 67.6 \\
    vs. MemoryTalker  & 63.8 & 68.5 \\
    vs. StreamingTalker  & 68.5 & 65.7 \\
    \bottomrule
  \end{tabular}
  \label{table:user_study_voca}
  \vspace{-20pt}
\end{wraptable}
Table~\ref{table:user_study_voca} demonstrates the results of a pairwise A/B preference study following~\cite{kim2025memorytalker}, comparing our method against seven competitors on \textbf{Lip Sync} and \textbf{Realism}.
We recruit 26 participants and exclude 2 who failed attention checks, resulting in 24 valid responses. 
Each participant rates 15 randomly sampled clips per competitor, with two videos shown side-by-side in randomized order for forced-choice decisions on each criterion.
Our method is preferred across all competitors, with average rates of 70.3\% (Lip Sync) and 70.7\% (Realism).
A one-sided binomial test against chance level (50\%) yields $p<0.0001$ for every competitor, confirming statistical significance.

\section{Conclusion}
In this paper, we propose a novel articulation-aware approach to speech-driven 3D facial animation grounded in visible articulation during human speech production.
We formulate speech-driven facial animation in terms of directional articulatory motions, together with Speech--Articulatory Memory (SAM) to model speech articulation correspondence and Topology-aware Articulatory Composition (TAC) to generate surface-consistent facial motion under mesh topology.
Experimental results demonstrate that the proposed formulation improves the speech consistency of facial animation.
By formulating facial motion through structured articulatory directions, our framework moves beyond holistic vertex regression toward physically grounded and interpretable speech-driven animation.

{
\small
\bibliographystyle{unsrt} 
\bibliography{neurips_2026}

@String{Computing = "Computing" }

@String{Computer = "{IEEE} Computer" }

@String{Springer = "Springer-Verlag" }

@article{rebernik2021review,
  title={A review of data collection practices using electromagnetic articulography},
  author={Rebernik, Teja and Jacobi, Jidde and Jonkers, Roel and Noiray, Aude and Wieling, Martijn},
  journal={Laboratory Phonology},
  volume={12},
  number={1},
  pages={6},
  year={2021},
  publisher={UBIQUITY PRESS LTD}
}

@article{anand2025teaching,
  title={Teaching Machines to Speak Using Articulatory Control},
  author={Anand, Akshay and Guo, Chenxu and Cho, Cheol Jun and Lian, Jiachen and Anumanchipalli, Gopala},
  journal={arXiv preprint arXiv:2510.05619},
  year={2025}
}

@incollection{maeda1990compensatory,
  title={Compensatory articulation during speech: Evidence from the analysis and synthesis of vocal-tract shapes using an articulatory model},
  author={Maeda, Shinji},
  booktitle={Speech production and speech modelling},
  pages={131--149},
  year={1990},
  publisher={Springer},
  address={Dordrecht}
}

@book{international1999handbook,
  title={Handbook of the International Phonetic Association: A guide to the use of the International Phonetic Alphabet},
  author={International Phonetic Association},
  year={1999},
  publisher={Cambridge University Press},
  address={Cambridge, UK} 
}

@inproceedings{kim25r_interspeech,
  title     = {{Learning Phonetic Context-Dependent Viseme for Enhancing Speech-Driven {3D} Facial Animation}},
  author    = {Hyung Kyu Kim and Hak Gu Kim},
  year      = {2025},
  booktitle = {{Interspeech 2025}},
  pages     = {3763--3767},
  doi       = {10.21437/Interspeech.2025-1692},
  issn      = {2958-1796},
}

@inproceedings{faceformer2022,
    title = {{FaceFormer}: Speech-Driven {3D} Facial Animation with Transformers},
    author = {Fan, Yingruo and Lin, Zhaojiang and Saito, Jun and Wang, Wenping and Komura, Taku},
    booktitle = {Proceedings of the IEEE/CVF Conference on Computer Vision and Pattern Recognition (CVPR)},
    year = {2022},
    pages = {18770--18780}, 
    publisher = {IEEE},
    address = {New Orleans, LA, USA}
}

@inproceedings{codetalker2023,
  title={{CodeTalker}: Speech-driven {3D} facial animation with discrete motion prior},
  author={Xing, Jinbo and Xia, Menghan and Zhang, Yuechen and Cun, Xiaodong and Wang, Jue and Wong, Tien-Tsin},
  booktitle={Proceedings of the IEEE/CVF Conference on Computer Vision and Pattern Recognition},
  pages={12780--12790},
  year={2023}
}

@inproceedings{selftalk23,
  title={{SelfTalk}: A Self-Supervised Commutative Training Diagram to Comprehend {3D} Talking Faces}, 
  author={Peng, Ziqiao and others},
  booktitle={Proceedings of the 31st ACM International Conference on Multimedia},
  pages = {5292--5301},
  year={2023},
  publisher={ACM},
  address={New York, NY, USA}
}

@InProceedings{imitator2023,
    author      = {Thambiraja, Balamurugan and Habibie, Ikhsanul and Aliakbarian, Sadegh and Cosker, Darren and Theobalt, Christian and Thies, Justus},
    title       = {Imitator: Personalized Speech-driven {3D} Facial Animation},
    booktitle   = {Proceedings of the IEEE/CVF International Conference on Computer Vision (ICCV)},
    month       = {October},
    year        = {2023},
    pages       = {20621-20631}
}

@ARTICLE{HuBERT2021,
    author      = {Hsu, Wei-Ning and Bolte, Benjamin and Tsai, Yao-Hung Hubert and Lakhotia, Kushal and Salakhutdinov, Ruslan and Mohamed, Abdelrahman},
    journal     = {IEEE/ACM Transactions on Audio, Speech, and Language Processing}, 
    title       = {{HuBERT}: Self-Supervised Speech Representation Learning by Masked Prediction of Hidden Units}, 
    year        = {2021},
    volume      = {29},
    number      = {},
    pages       = {3451-3460},
    doi         = {10.1109/TASLP.2021.3122291}
}

@INPROCEEDINGS{VOCA2019,
    author={Cudeiro, Daniel and Bolkart, Timo and Laidlaw, Cassidy and Ranjan, Anurag and Black, Michael J.},
    booktitle={2019 IEEE/CVF Conference on Computer Vision and Pattern Recognition (CVPR)}, 
    title={Capture, Learning, and Synthesis of {3D} Speaking Styles}, 
    year={2019},
    volume={},
    number={},
    pages={10093-10103},
    publisher={IEEE},
    doi={10.1109/CVPR.2019.01034}
}

@article{BIWI2010,
  title={A 3-d audio-visual corpus of affective communication},
  author={Fanelli, Gabriele and Gall, Juergen and Romsdorfer, Harald and Weise, Thibaut and Van Gool, Luc},
  journal={IEEE Transactions on Multimedia},
  volume={12},
  number={6},
  pages={591--598},
  year={2010},
  publisher={IEEE}
}

@article{FLAME2017,
author = {Li, Tianye and Bolkart, Timo and Black, Michael J. and Li, Hao and Romero, Javier},
title = {Learning a model of facial shape and expression from 4D scans},
year = {2017},
issue_date = {December 2017},
publisher = {Association for Computing Machinery},
address = {New York, NY, USA},
volume = {36},
number = {6},
issn = {0730-0301},
url = {https://doi.org/10.1145/3130800.3130813},
doi = {10.1145/3130800.3130813},
journal = {ACM Trans. Graph.},
month = nov,
articleno = {194},
numpages = {17}
}

@INPROCEEDINGS{meshtalk2021,
    author={Richard, Alexander and Zollhöfer, Michael and Wen, Yandong and de la Torre, Fernando and Sheikh, Yaser},
    booktitle={IEEE/CVF International Conference on Computer Vision (ICCV)}, 
    title={{MeshTalk}: {3D} Face Animation from Speech using Cross-Modality Disentanglement}, 
    year={2021},
    publisher={IEEE},
    pages={1153-1162},
    doi={10.1109/ICCV48922.2021.00121}}

@inproceedings{mimic2024,
    title={Mimic: Speaking Style Disentanglement for Speech-Driven {3D} Facial Animation},
  author = {{Hui Fu} and {Zeqing Wang} and {Ke Gong} and {Keze Wang} and {Tianshui Chen} and {Haojie Li} and {Haifeng Zeng} and {Wenxiong Kang}},
    booktitle={The 38th Annual AAAI Conference on Artificial Intelligence (AAAI)},
    year={2024}
}

@inproceedings{CompositeandRegionalFacialMovements,
author = {Wu, Haozhe and Zhou, Songtao and Jia, Jia and Xing, Junliang and Wen, Qi and Wen, Xiang},
title = {Speech-Driven {3D} Face Animation with Composite and Regional Facial Movements},
year = {2023},
isbn = {9798400701085},
publisher = {Association for Computing Machinery},
address = {New York, NY, USA},
url = {https://doi.org/10.1145/3581783.3611775},
doi = {10.1145/3581783.3611775},
booktitle = {Proceedings of the 31st ACM International Conference on Multimedia},
pages = {6822–6830},
numpages = {9},
location = {Ottawa ON, Canada},
series = {MM '23}
}

@InProceedings{emotalk23,
    author    = {Peng, Ziqiao and Wu, Haoyu and Song, Zhenbo and Xu, Hao and Zhu, Xiangyu and He, Jun and Liu, Hongyan and Fan, Zhaoxin},
    title     = {{EmoTalk}: Speech-Driven Emotional Disentanglement for {3D} Face Animation},
    booktitle = {Proceedings of the IEEE/CVF International Conference on Computer Vision (ICCV)},
    month     = {October},
    year      = {2023},
    pages     = {20687-20697}
}

@article{dtw07,
  title={Toward accurate dynamic time warping in linear time and space},
  author={Salvador, Stan and Chan, Philip},
  journal={Intelligent Data Analysis},
  volume={11},
  number={5},
  pages={561--580},
  year={2007},
  publisher={IOS Press}
}

@inproceedings{icip2024,
  author={Kim, Hyung Kyu and Lee, Sangmin and Kim, Hak Gu},
  booktitle={2024 IEEE International Conference on Image Processing (ICIP)}, 
  title={Analyzing Visible Articulatory Movements in Speech Production For Speech-Driven {3D} Facial Animation}, 
  year={2024},
    publisher={IEEE},
  volume={},
  number={},
  pages={3575-3579}
}

@inproceedings{facetalk2024,
  title={FaceTalk: Audio-Driven Motion Diffusion for Neural Parametric Head Models},
  author={Aneja, Shivangi and Thies, Justus and Dai, Angela and Niessner, Matthias},
  booktitle={Proceedings of the IEEE/CVF Conference on Computer Vision and Pattern Recognition (CVPR)},
  pages={21263--21273},
  year={2024},
  publisher={IEEE},
  address={Seattle, WA, USA},
  doi={10.1109/CVPR52733.2024.02009},
  url={https://ieeexplore.ieee.org/document/10658217}
}

@article{diffposetalk2024,
  title={{Diffposetalk}: Speech-driven stylistic {3D} facial animation and head pose generation via diffusion models},
  author={Sun, Zhiyao and Lv, Tian and Ye, Sheng and Lin, Matthieu and Sheng, Jenny and Wen, Yu-Hui and Yu, Minjing and Liu, Yong-jin},
  journal={ACM Transactions on Graphics (TOG)},
  volume={43},
  number={4},
  pages={1--9},
  year={2024},
  publisher={ACM New York, NY, USA}
}

@inproceedings{kmtalk2024,
    title={KMTalk: Speech-Driven 3D Facial Animation with Key Motion Embedding},
    author={Xu, Zhihao and Gong, Shengjie and Tang, Jiapeng and Liang, Lingyu and Huang, Yining and Li, Haojie and Huang, Shuangping},
    booktitle={European Conference on Computer Vision},
    pages={236--253},
    year={2024},
    organization={Springer}
}

@inproceedings{scantalk2024,
    title={Scantalk: 3d talking heads from unregistered scans},
    author={Nocentini, Federico and Besnier, Thomas and Ferrari, Claudio and Arguillere, Sylvain and Berretti, Stefano and Daoudi, Mohamed},
    booktitle={European Conference on Computer Vision},
    pages={19--36},
    year={2024},
    organization={Springer}
}

@inproceedings{unitalker2024,
  title={UniTalker: Scaling up Audio-Driven 3D Facial Animation through A Unified Model},
  author={Fan, Xiangyu and Li, Jiaqi and Lin, Zhiqian and Xiao, Weiye and Yang, Lei},
  booktitle={European Conference on Computer Vision (ECCV)},
  pages={204--221},
  year={2024},
  publisher={Springer},
  address={Cham}
}

@ARTICLE{zhuang2024learn2talk,
    author={Zhuang, Yixiang and Cheng, Baoping and Cheng, Yao and Jin, Yuntao and Liu, Renshuai and Li, Chengyang and Cheng, Xuan and Liao, Jing and Lin, Juncong},
    journal={IEEE Transactions on Visualization and Computer Graphics}, 
    title={Learn2Talk: 3D Talking Face Learns from 2D Talking Face}, 
    year={2024},
    volume={},
    number={},
    pages={1-13},
    doi={10.1109/TVCG.2024.3476275}
}

@inproceedings{chatziagapi2023avface,
  title={AVFace: Towards Detailed Audio-Visual 4D Face Reconstruction},
  author={Chatziagapi, Aggelina and Samaras, Dimitris},
  booktitle={Proceedings of the IEEE/CVF Conference on Computer Vision and Pattern Recognition (CVPR)},
  pages={16878--16889},
  year={2023},
  publisher={IEEE},
  address={Vancouver, BC, Canada},
  doi={10.1109/CVPR52729.2023.01619},
  url={https://openaccess.thecvf.com/content/CVPR2023/html/Chatziagapi_AVFace_Towards_Detailed_Audio-Visual_4D_Face_Reconstruction_CVPR_2023_paper.html}
}

@inproceedings{he2023speech4mesh,
  title={Speech4mesh: Speech-assisted monocular 3d facial reconstruction for speech-driven 3d facial animation},
  author={He, Shan and He, Haonan and Yang, Shuo and Wu, Xiaoyan and Xia, Pengcheng and Yin, Bing and Liu, Cong and Dai, Lirong and Xu, Chang},
  booktitle={Proceedings of the IEEE/CVF International Conference on Computer Vision},
  pages={14192--14202},
  year={2023}
}

@inproceedings{yang2023semi,
  title={Semi-supervised Speech-driven 3D Facial Animation via Cross-modal Encoding},
  author={Yang, Peiji and Wei, Huawei and Zhong, Yicheng and Wang, Zhisheng},
  booktitle={Proceedings of the IEEE/CVF International Conference on Computer Vision},
  pages={21032--21041},
  year={2023}
}

@article{lofqvist2005lip,
  title={Lip kinematics in long and short stop and fricative consonants},
  author={L{\"o}fqvist, Anders},
  journal={The Journal of the Acoustical Society of America},
  volume={117},
  number={2},
  pages={858--878},
  year={2005},
  publisher={ASA},
  doi={10.1121/1.1840531}
}

@article{wang2013articulatory,
  title={Articulatory distinctiveness of vowels and consonants: a data-driven approach},
  author={Wang, Jun and Green, Jordan R and Samal, Ashok and Yunusova, Yana},
  journal={Journal of Speech, Language, and Hearing Research},
  volume={56},
  number={5},
  pages={1539--1551},
  year={2013},
  doi={10.1044/1092-4388(2013/12-0030)}
}

@inproceedings{wav2lip,
author = {Prajwal, K R and Mukhopadhyay, Rudrabha and Namboodiri, Vinay P. and Jawahar, C.V.},
title = {A Lip Sync Expert Is All You Need for Speech to Lip Generation In the Wild},
year = {2020},
isbn = {9781450379885},
publisher = {Association for Computing Machinery},
address = {New York, NY, USA},
url = {https://doi.org/10.1145/3394171.3413532},
doi = {10.1145/3394171.3413532},
booktitle = {Proceedings of the 28th ACM International Conference on Multimedia},
pages = {484–492},
numpages = {9},
location = {Seattle, WA, USA},
series = {MM '20}
}

@article{craig2008linear,
  title={A linear model of acoustic-to-facial mapping: Model parameters, data set size, and generalization across speakers},
  author={Craig, Matthew S and Van Lieshout, Pascal and Wong, Willy},
  journal={The Journal of the Acoustical Society of America},
  volume={124},
  number={5},
  pages={3183--3190},
  year={2008},
  publisher={AIP Publishing}
}

@article{lucero2008analysis,
  title={Analysis of facial motion patterns during speech using a matrix factorization algorithm},
  author={Lucero, Jorge C and Munhall, Kevin G},
  journal={The Journal of the Acoustical Society of America},
  volume={124},
  number={4},
  pages={2283--2290},
  year={2008},
  publisher={AIP Publishing}
}

@inproceedings{kim2025memorytalker,
  title={{MemoryTalker}: Personalized Speech-Driven 3D Facial Animation via Audio-Guided Stylization},
  author={Kim, Hyung Kyu and Lee, Sangmin and Kim, Hak Gu},
  booktitle={Proceedings of the IEEE/CVF International Conference on Computer Vision},
  pages={11241--11251},
  year={2025}
}

@inproceedings{FaceDiffuser23,
author = {Stan, Stefan and Haque,  Kazi Injamamul and Yumak,  Zerrin},
title = {{FaceDiffuser}: Speech-Driven 3D Facial Animation Synthesis Using Diffusion},
booktitle = {ACM SIGGRAPH Conference on Motion, Interaction and Games (MIG '23), November 15--17, 2023, Rennes, France},
year = {2023},
location = {Rennes, France},
numpages = {11},
url = {https://doi.org/10.1145/3623264.3624447},
doi = {10.1145/3623264.3624447},
publisher = {ACM},
address = {New York, NY, USA},
}

@inproceedings{thambiraja2025diface,
title={3Di{FACE}: Synthesizing and Editing Holistic 3D Facial Animation},
author={Balamurugan Thambiraja and Malte Prinzler and Sadegh Aliakbarian and Darren Cosker and Justus Thies},
booktitle={International Conference on 3D Vision 2025},
year={2025},
url={https://openreview.net/forum?id=8qpjYG1x8I}
}

@inproceedings{chen2025diffusiontalker,
  title={{DiffusionTalker}: Efficient and Compact Speech-Driven 3D Talking Head via Personalizer-Guided Distillation},
  author={Chen, Peng and Wei, Xiaobao and Lu, Ming and Chen, Hui and Tian, Feng},
  booktitle={Proceedings of the IEEE International Conference on Multimedia and Expo (ICME)},
  year={2025},
  publisher={IEEE},
  address={New York, NY, USA}
}

@inproceedings{chu2025artalkspeechdriven3dhead,
author = {Chu, Xuangeng and Goswami, Nabarun and Cui, Ziteng and Wang, Hanqin and Harada, Tatsuya},
title = {{ARTalk}: Speech-Driven 3D Head Animation via Autoregressive Model},
year = {2025},
isbn = {9798400721373},
publisher = {Association for Computing Machinery},
address = {New York, NY, USA},
url = {https://doi.org/10.1145/3757377.3763955},
doi = {10.1145/3757377.3763955},
booktitle = {Proceedings of the SIGGRAPH Asia 2025 Conference Papers},
articleno = {55},
numpages = {9},
location = {
},
series = {SA Conference Papers '25}
}

@inproceedings{li2025wav2sem,
  title={{Wav2Sem}: Plug-and-Play Audio Semantic Decoupling for 3D Speech-Driven Facial Animation},
  author={Li, Hao and Dai, Ju and Zhao, Xin and Zhou, Feng and Pan, Junjun and Li, Lei},
  booktitle={Proceedings of the Computer Vision and Pattern Recognition Conference},
  pages={183--192},
  year={2025}
}

@inproceedings{pan2025model,
  title={Model See Model Do: Speech-Driven Facial Animation with Style Control},
  author={Pan, Yifang and Singh, Karan and Hafemann, Luiz Gustavo},
  booktitle={Proceedings of the Special Interest Group on Computer Graphics and Interactive Techniques Conference Conference Papers},
  pages={1--10},
  year={2025}
}

@inproceedings{xieecoface,
  title={{EcoFace}: Audio-Visual Emotional Co-Disentanglement Speech-Driven 3D Talking Face Generation},
  author={Xie, Jiajian and Zhang, Shengyu and Li, Mengze and Zhao, Zhou and Wu, Fei and others},
  booktitle={The Thirteenth International Conference on Learning Representations},
    year={2025}
}

@misc{wuu2023multifacedatasetneuralface,
  title={{Multiface}: A dataset for neural face rendering},
  author={Wuu, Cheng-hsin and Zheng, Ningyuan and Ardisson, Scott and Bali, Rohan and Belko, Danielle and Brockmeyer, Eric and Evans, Lucas and Godisart, Timothy and Ha, Hyowon and Huang, Xuhua and others},
  journal={arXiv preprint arXiv:2207.11243},
  year={2022}
}

@inproceedings{georges22_interspeech,
  title     = {Self-supervised speech unit discovery from articulatory and acoustic features using VQ-VAE},
  author    = {Marc-Antoine Georges and Jean-Luc Schwartz and Thomas Hueber},
  year      = {2022},
  booktitle = {Interspeech 2022},
  pages     = {774--778},
  doi       = {10.21437/Interspeech.2022-10876},
  issn      = {2958-1796},
}

@inproceedings{chung24_interspeech,
  title     = {Speaker-Independent Acoustic-to-Articulatory Inversion through Multi-Channel Attention Discriminator},
  author    = {Woo-Jin Chung and Hong-Goo Kang},
  year      = {2024},
  booktitle = {Interspeech 2024},
  pages     = {1540--1544},
  doi       = {10.21437/Interspeech.2024-1269},
  issn      = {2958-1796},
}

@inproceedings{sun22b_interspeech,
  title     = {Unsupervised Acoustic-to-Articulatory Inversion with Variable Vocal Tract Anatomy},
  author    = {Yifan Sun and Qinlong Huang and Xihong Wu},
  year      = {2022},
  booktitle = {Interspeech 2022},
  pages     = {4656--4660},
  doi       = {10.21437/Interspeech.2022-477},
  issn      = {2958-1796},
}

@inproceedings{kim2025deeptalk,
  title={{DEEPTalk}: Dynamic Emotion Embedding for Probabilistic Speech-Driven 3D Face Animation},
  author={Kim, Jisoo and others},
  booktitle={Proceedings of the AAAI Conference on Artificial Intelligence},
  pages={4275--4283},
  year={2025},
  publisher={AAAI Press},
  address={Washington, DC, USA}
}

@article{tanaka2022acceptability,
  title={The acceptability of virtual characters as social skills trainers: usability study},
  author={Tanaka, Hiroki and Nakamura, Satoshi},
  journal={JMIR human factors},
  volume={9},
  number={1},
  pages={e35358},
  year={2022},
  publisher={JMIR Publications Inc., Toronto, Canada}
}

@inproceedings{zhang2021flow,
  title={Flow-Guided One-Shot Talking Face Generation With a High-Resolution Audio-Visual Dataset},
  author={Zhang, Zhimeng and Li, Lincheng and Ding, Yu and Fan, Changjie},
  booktitle={Proceedings of the IEEE/CVF Conference on Computer Vision and Pattern Recognition},
  pages={3661--3670},
  year={2021}
}

@inproceedings{parrell18_interspeech,
  title     = {FACTS: A Hierarchical Task-based Control Model of Speech Incorporating Sensory Feedback},
  author    = {Benjamin Parrell and Vikram Ramanarayanan and Srikantan Nagarajan and John Houde},
  year      = {2018},
  booktitle = {Interspeech 2018},
  pages     = {1497--1501},
  doi       = {10.21437/Interspeech.2018-2087},
  issn      = {2958-1796},
}

@inproceedings{FaceXHuBERT_ICMI23,
    author = {Haque,  Kazi Injamamul and Yumak,  Zerrin},
    title = {{FaceXHuBERT}: Text-less Speech-driven {E(X)}pressive 3D Facial Animation Synthesis Using Self-Supervised Speech Representation Learning},
    booktitle = {International Conference on Multimodal Interaction (ICMI ’23)},
    year = {2023},
    location = {Paris, France},
    numpages = {10},
    url = {https://doi.org/10.1145/3577190.3614157},
    doi = {10.1145/3577190.3614157},
    publisher = {ACM},
    address = {New York, NY, USA},
}

@article{zhou2018visemenet,
  title={{VisemeNet}: {Audio-Driven Animator-Centric Speech Animation}},
  author={Zhou, Yang and Xu, Zhan and Landreth, Chris and Kalogerakis, Evangelos and Maji, Subhransu and Singh, Karan},
  journal={ACM Transactions on Graphics (ToG)},
  volume={37},
  number={4},
  pages={1--10},
  year={2018},
  publisher={ACM New York, NY, USA}
}

@article{edwards2016jali,
  title={Jali: an animator-centric viseme model for expressive lip synchronization},
  author={Edwards, Pif and Landreth, Chris and Fiume, Eugene and Singh, Karan},
  journal={ACM Transactions on graphics (TOG)},
  volume={35},
  number={4},
  pages={1--11},
  year={2016},
  publisher={ACM New York, NY, USA}
}

@inproceedings{qu2025exptalk,
  title={ExpTalk: Diverse Emotional Expression via Adaptive Disentanglement and Refined Alignment for Speech-Driven 3D Facial Animation},
  author={Qu, Zhan and Zhang, Shengyu and Li, Mengze and Chen, Zhuo and Lv, Chengfei and Zhao, Zhou and Wu, Fei},
  booktitle={Proceedings of the Thirty-Fourth International Joint Conference on Artificial Intelligence},
  pages={1811--1819},
  year={2025}
}

@inproceedings{danvevcek2023emotional,
  title={Emotional speech-driven animation with content-emotion disentanglement},
  author={Dan{\v{e}}{\v{c}}ek, Radek and Chhatre, Kiran and Tripathi, Shashank and Wen, Yandong and Black, Michael and Bolkart, Timo},
  booktitle={SIGGRAPH Asia 2023 Conference Papers},
  pages={1--13},
  year={2023}
}

@article{lin2024emoface,
  title={{EmoFace}: Emotion-Content Disentangled Speech-Driven 3D Talking Face Animation},
  author={Lin, Yihong and Peng, Liang and Fan, Zhaoxin and Wu, Xianjia and Hu, Jianqiao and Li, Xiandong and Kang, Wenxiong and Lei, Songju},
  journal={arXiv preprint arXiv:2408.11518},
  year={2024}
}

@inproceedings{wu2024probtalk3d,
  title={{Probtalk3d}: Non-deterministic emotion controllable speech-driven 3d facial animation synthesis using vq-vae},
  author={Wu, Sichun and Haque, Kazi Injamamul and Yumak, Zerrin},
  booktitle={Proceedings of the 17th ACM SIGGRAPH conference on motion, interaction, and games},
  pages={1--12},
  year={2024}
}

@inproceedings{yang2024probabilistic,
  title={Probabilistic speech-driven 3d facial motion synthesis: New benchmarks methods and applications},
  author={Yang, Karren D and Ranjan, Anurag and Chang, Jen-Hao Rick and Vemulapalli, Raviteja and Tuzel, Oncel},
  booktitle={Proceedings of the IEEE/CVF conference on computer vision and pattern recognition},
  pages={27294--27303},
  year={2024}
}

@inproceedings{kipf2017semi,
  title={Semi-Supervised Classification with Graph Convolutional Networks},
  author={Kipf, Thomas N. and Welling, Max},
  booktitle={International Conference on Learning Representations (ICLR)},
  year={2017}
}

@inproceedings{yang2026streamingtalker,
  title={StreamingTalker: Audio-driven 3D Facial Animation with Autoregressive Diffusion Model},
  author={Yang, Yifan and Cen, Zhi and Peng, Sida and Chen, Xiangwei and Deng, Yifu and Zhu, Xinyu and Jia, Fan and Zhou, Xiaowei and Bao, Hujun},
  booktitle={Proceedings of the AAAI Conference on Artificial Intelligence},
  volume={40},
  number={14},
  pages={11766--11774},
  year={2026}
}

@article{atal1978inversion,
  title={Inversion of articulatory-to-acoustic transformation in the vocal tract by a computer-sorting technique},
  author={Atal, Bishnu S and Chang, Jih Jie and Mathews, Max V and Tukey, John W},
  journal={The Journal of the Acoustical Society of America},
  volume={63},
  number={5},
  pages={1535--1555},
  year={1978},
  publisher={Acoustical Society of America}
}

@article{panchapagesan2011study,
  title={A study of acoustic-to-articulatory inversion of speech by analysis-by-synthesis using chain matrices and the Maeda articulatory model},
  author={Panchapagesan, Sankaran and Alwan, Abeer},
  journal={The Journal of the Acoustical Society of America},
  volume={129},
  number={4},
  pages={2144--2162},
  year={2011},
  publisher={AIP Publishing}
}

@article{ghosh2010generalized,
  title={A generalized smoothness criterion for acoustic-to-articulatory inversion},
  author={Ghosh, Prasanta Kumar and Narayanan, Shrikanth},
  journal={The Journal of the Acoustical Society of America},
  volume={128},
  number={4},
  pages={2162--2172},
  year={2010},
  publisher={AIP Publishing}
}

@article{haque2025probtalk3d,
  title={{ProbTalk3D-X}: Prosody enhanced non-deterministic emotion controllable speech-driven 3D facial animation synthesis},
  author={Haque, Kazi Injamamul and Wu, Sichun and Yumak, Zerrin},
  journal={Computers \& Graphics},
  pages={104358},
  year={2025},
  publisher={Elsevier}
}

@article{badin2002three,
  title={Three-dimensional linear articulatory modeling of tongue, lips and face, based on MRI and video images},
  author={Badin, Pierre and Bailly, Gerard and Reveret, Lionel and Baciu, Monica and Segebarth, Christoph and Savariaux, Christophe},
  journal={Journal of Phonetics},
  volume={30},
  number={3},
  pages={533--553},
  year={2002},
  publisher={Elsevier}
}

@inproceedings{shen2024deitalk,
  title={{Deitalk}: Speech-driven 3d facial animation with dynamic emotional intensity modeling},
  author={Shen, Kang and Xia, Haifeng and Geng, Guangxing and Geng, Guangyue and Xia, Siyu and Ding, Zhengming},
  booktitle={Proceedings of the 32nd ACM international conference on multimedia},
  pages={10506--10514},
  year={2024}
}

@article{jiang2026editemotalk,
  title={{EditEmoTalk}: Controllable Speech-Driven 3D Facial Animation with Continuous Expression Editing},
  author={Jiang, Diqiong and Zhu, Kai and Song, Dan and Chang, Jian and Chen, Chenglizhao and Wu, Zhenyu},
  journal={arXiv preprint arXiv:2601.10000},
  year={2026}
}

@article{chu2025dcptalk,
  title={{Dcptalk}: Speech-driven {3D} face animation with personalized facial dynamic coupling properties},
  author={Chu, Zhaojie and Guo, Kailing and Xing, Xiaofen and Liu, Pengsheng and Cai, Bolun and Xu, Xiangmin},
  journal={IEEE Transactions on Multimedia},
  volume={27},
  pages={4427--4440},
  year={2025},
  publisher={IEEE}
}

@inproceedings{10.1145/3757377.3763887,
author = {Lu, Xin and Zhuang, Chuanqing and Jin, Chenxi and Lu, Zhengda and Wang, Yiqun and Liu, Wu and Xiao, Jun},
title = {{LSF-Animation}: Label-Free Speech-Driven Facial Animation via Implicit Feature Representation},
year = {2025},
isbn = {9798400721373},
publisher = {Association for Computing Machinery},
address = {New York, NY, USA},
url = {https://doi.org/10.1145/3757377.3763887},
doi = {10.1145/3757377.3763887},
booktitle = {Proceedings of the SIGGRAPH Asia 2025 Conference Papers},
articleno = {52},
numpages = {12},
location = {
},
series = {SA Conference Papers '25}
}

@article{chu2026siatalker,
  title={{SiaTalker}: Siamese Emotion Injection for Coarse-to-Fine Speech-Driven {3D} Facial Animation},
  author={Chu, Zhaojie and Lan, Yilin and Jin, Jianxiu and Xu, Xiangmin and Xing, Xiaofen},
  journal={IEEE Transactions on Affective Computing},
  year={2026},
  publisher={IEEE}
}

@article{chu2026eetalk,
  title={{EETalk}: Expression Enhancement in Speech-Driven {3D} Facial Animation},
  author={Chu, Zhaojie and Guo, Kailing and Xing, Xiaofen and Cai, Bolun and Wang, Lin and Xu, Xiangmin},
  journal={IEEE Transactions on Multimedia},
  year={2026},
  publisher={IEEE}
}

@article{zhuang2026talkingeyes,
  title={{TalkingEyes}: Pluralistic speech-driven {3D} eye gaze animation},
  author={Zhuang, Yixiang and Ma, Chunshan and Cheng, Yao and Cheng, Xuan and Liao, Jing and Lin, Juncong},
  journal={IEEE Transactions on Visualization and Computer Graphics},
  year={2026},
  publisher={IEEE}
}

@article{zhang2026ex,
  title={{Ex-Omni}: Enabling 3D Facial Animation Generation for Omni-modal Large Language Models},
  author={Zhang, Haoyu and Li, Zhipeng and Guo, Yiwen and Yu, Tianshu},
  journal={arXiv preprint arXiv:2602.07106},
  year={2026}
}

@article{chartier2018encoding,
  author  = {Chartier, Josh and Anumanchipalli, Gopala K. and Johnson, Keith and Chang, Edward F.},
  title   = {Encoding of Articulatory Kinematic Trajectories in Human Speech Sensorimotor Cortex},
  journal = {Neuron},
  volume  = {98},
  number  = {5},
  pages   = {1042--1054},
  year    = {2018},
}

@article{DiffusionNet,
author = {Sharp, Nicholas and Attaiki, Souhaib and Crane, Keenan and Ovsjanikov, Maks},
title = {{DiffusionNet}: Discretization Agnostic Learning on Surfaces},
year = {2022},
issue_date = {June 2022},
publisher = {Association for Computing Machinery},
address = {New York, NY, USA},
volume = {41},
number = {3},
issn = {0730-0301},
url = {https://doi.org/10.1145/3507905},
doi = {10.1145/3507905},
journal = {ACM Trans. Graph.},
month = mar,
articleno = {27},
numpages = {16}
}

@article{savitzky1964smoothing,
  title={Smoothing and differentiation of data by simplified least squares procedures.},
  author={Savitzky, Abraham and Golay, Marcel JE},
  journal={Analytical chemistry},
  volume={36},
  number={8},
  pages={1627--1639},
  year={1964},
  publisher={ACS Publications}
}

@article{fromkin1964lip,
  title={Lip positions in American English vowels},
  author={Fromkin, Victoria},
  journal={Language and speech},
  volume={7},
  number={4},
  pages={215--225},
  year={1964},
  publisher={SAGE Publications Sage UK: London, England}
}

@article{montgomery1983physical,
  title={Physical characteristics of the lips underlying vowel lipreading performance},
  author={Montgomery, Allen A and Jackson, Pamela L},
  journal={The Journal of the Acoustical Society of America},
  volume={73},
  number={6},
  pages={2134--2144},
  year={1983},
  publisher={Acoustical Society of America}
}
}

\newpage
\appendix
\section{Appendix}

\renewcommand{\thefigure}{S\arabic{figure}}
\renewcommand{\thetable}{S\arabic{table}}
\setcounter{figure}{0}
\setcounter{table}{0}
\renewcommand{\theequation}{S\arabic{equation}}
\setcounter{equation}{0}

\subsection{Additional Implementation and Evaluation Details}
\label{bc5}

\subsubsection{Implementation Details}
\label{supp:Implementation_Details}

\paragraph{\textbf{Motion Autoencoder -- Architecture Details.}}
\label{supp:archi_detail}
The motion autoencoder used for $\mathcal{L}_{\mathrm{store}}$ consists of three independent encoder--decoder pairs, one per articulatory direction $\delta \in \{S, O, P\}$. Each pair operates on the corresponding directional component of the ground-truth motion $\mathbf{v}_\delta^t \in \mathbb{R}^{V}$, encoding it into the latent motion feature $\mathbf{z}_\delta^t \in \mathbb{R}^{D_{\mathrm{SAM}}}$ via the Transformer encoder and reconstructing $\mathbf{v}_\delta^t$ from $\mathbf{z}_\delta^t$ via the Transformer decoder. The encoder output $\mathbf{z}_\delta^t$ serves as the alignment target for the retrieved feature $\mathbf{f}_{\mathrm{retri},\delta}^t$ in Eq.~\ref{main:store_loss}. The per-direction structure is summarized in Table~\ref{tab:ae_arch}.

\begin{table}[h]
\centering
\small
\setlength{\tabcolsep}{5pt}
\renewcommand{\arraystretch}{1.2}
\caption{Per-direction architecture of the motion autoencoder used to produce $\mathbf{z}_\delta^t$. Three encoder--decoder pairs are jointly trained, one for each $\delta \in \{S, O, P\}$.}
\vspace{5pt}
\begin{tabular}{llc}
\toprule
Stage & Component & Output shape \\
\midrule
Input        & $\mathbf{v}_\delta^{1:T}$ (per-direction motion)       & $(T, V)$ \\
\midrule
\multirow{3}{*}{Encoder}
             & Input projection ($V \to D_{\mathrm{SAM}}$, Linear)     & $(T, D_{\mathrm{SAM}})$ \\
             & Sinusoidal positional encoding                           & $(T, D_{\mathrm{SAM}})$ \\
             & Transformer encoder (1 layer, 4 heads, FFN $4{\times}D_{\mathrm{SAM}}$, GELU, dropout 0.1) & $(T, D_{\mathrm{SAM}})$ \\
\midrule
Latent       & $\mathbf{z}_\delta^{1:T}$ (motion feature)              & $(T, D_{\mathrm{SAM}})$ \\
\midrule
\multirow{2}{*}{Decoder}
             & Transformer decoder (1 layer, 4 heads, FFN $4{\times}D_{\mathrm{SAM}}$, GELU, dropout 0.1) & $(T, D_{\mathrm{SAM}})$ \\
             & Output projection ($D_{\mathrm{SAM}} \to V$, Linear)    & $(T, V)$ \\
\bottomrule
\end{tabular}
\label{tab:ae_arch}
\end{table}

\paragraph{\textbf{Motion Autoencoder -- Training Details.}}
For each dataset (VOCASET~\cite{VOCA2019} and TFHP~\cite{diffposetalk2024}), we pre-train the three directional autoencoders before freezing them for SAM training. For each sequence, the per-direction motion $\mathbf{v}_\delta^{1:T}$ is normalized along the time axis via per-vertex $z$-score statistics, passed through the encoder--decoder pair, and unnormalized back to the raw vertex domain. We minimize the sum of a reconstruction loss and a velocity loss in the raw vertex domain:
\begin{equation}
\mathcal{L}_{\mathrm{AE}} = \sum_{\delta \in \{S,O,P\}}
\left( \mathcal{L}_{\mathrm{recon},\delta}^{\mathrm{AE}} + \mathcal{L}_{\mathrm{vel},\delta}^{\mathrm{AE}} \right),
\end{equation}
where $\mathcal{L}_{\mathrm{recon},\delta}^{\mathrm{AE}}$ and $\mathcal{L}_{\mathrm{vel},\delta}^{\mathrm{AE}}$ are the MSE between the predicted and ground-truth motion and their first-order temporal differences, respectively. We train for up to 300 epochs with Adam (learning rate $10^{-4}$, batch size 1) on a single NVIDIA RTX 4090, with $D_{\mathrm{SAM}}{=}128$. After pre-training, all autoencoder parameters are frozen, and only the encoder output $\mathbf{z}_\delta^t$ is used to supervise the value memory via $\mathcal{L}_{\mathrm{store}}$.

\paragraph{\textbf{SAM -- Periodic Positional Encoding.}}
\label{supp:ppe}
Periodic positional encoding (PPE)~\cite{faceformer2022} injects time-step information into transformer inputs while encoding temporal periodicity. 
Unlike standard sinusoidal positional encoding, which assigns a unique embedding to every position, PPE resets after a fixed period $\rho$ and repeats the same set of $\rho$ embeddings across the sequence, providing cyclic temporal cues aligned with the rhythmic nature of speech.

Specifically, $\mathbf{P} \in \mathbb{R}^{T \times D_{\mathrm{SAM}}}$ is defined for row index $t \in \{1, \dots, T\}$ and column index $d \in \{0, \dots, D_{\mathrm{SAM}}-1\}$ with $k = \lfloor d/2 \rfloor$:
\begin{equation}
\mathbf{P}_{t, 2k} = \sin\!\left(\frac{t \bmod \rho}{10000^{2k/D_{\mathrm{SAM}}}}\right), \quad
\mathbf{P}_{t, 2k+1} = \cos\!\left(\frac{t \bmod \rho}{10000^{2k/D_{\mathrm{SAM}}}}\right).
\end{equation}
Here $t \bmod \rho$ denotes the remainder when $t$ is divided by $\rho$ (e.g., $26 \bmod 25 = 1$), so the positional encoding cycles every $\rho$ time steps.
We set $\rho$ to match the frame rate of each dataset so that one positional cycle spans approximately one second of speech: $\rho = 30$ for VOCASET (30~fps) and $\rho = 25$ for TFHP (25~fps), following~\cite{faceformer2022}.

\subsubsection{Evaluation Details}
\label{supp:evaluation_details}

\paragraph{\textbf{Articulatory Tendency -- Computation Details.}}
\label{supp:tendency_analysis}
We detail the computation used for the articulatory tendency analysis shown in Fig.~\ref{fig3:radar_plot}.
For each frame $t$, we compute three signed 1D trajectories from the four lip anchors (UL, LL, LMC, RMC), each referenced to the temporal median of its raw signal:
\begin{equation}
\label{supp:trajectories}
    \begin{aligned}
        s(t) &= (\mathbf{p}_{\mathrm{LMC},x}(t) - \mathbf{p}_{\mathrm{RMC},x}(t)) - (\bar{\mathbf{p}}_{\mathrm{LMC},x} - \bar{\mathbf{p}}_{\mathrm{RMC},x}), \\
        o(t) &= (\mathbf{p}_{\mathrm{UL},y}(t) - \mathbf{p}_{\mathrm{LL},y}(t)) - (\bar{\mathbf{p}}_{\mathrm{UL},y} - \bar{\mathbf{p}}_{\mathrm{LL},y}), \\
        p(t) &= (\mathbf{p}_{\mathrm{UL},z}(t) - \bar{\mathbf{p}}_{\mathrm{UL},z}) + (\mathbf{p}_{\mathrm{LL},z}(t) - \bar{\mathbf{p}}_{\mathrm{LL},z}),
    \end{aligned}
\end{equation}
where $s(t)$, $o(t)$, $p(t)$ correspond to the spreading, opening, and protrusion signals, and $\bar{\mathbf{p}}_*$ denotes the temporal median of the corresponding signal across all frames.
Unlike the absolute distance signals in Sec.~4.3, these are kept signed so that the slope sign can recover the direction of motion (e.g., opening vs.\ closing).

We compute a per-frame slope by a centered local-slope estimator over a 7-frame window ($K{=}3$), which corresponds to the Savitzky--Golay 1st-order derivative filter~\cite{savitzky1964smoothing}:
\begin{equation}
    \dot{y}(t) = \frac{\sum_{i=-K}^{K} i \cdot y(t+i)}{\sum_{i=-K}^{K} i^2},
\end{equation}
where $y \in \{s, o, p\}$ and boundaries use reflective padding.
This is the slope of the linear fit through the $2K+1$ samples centered at $t$, providing a smoother trend estimate than a single-frame finite difference.

We split each slope by sign into paired non-negative components, where positive slope corresponds to motion in the canonical direction (spreading, opening, protrusion) and negative slope to its opposite (rounding, closing, retraction):
\begin{equation}
    \begin{aligned}
        \mathrm{Sp}(t) &= \max(\dot{s}(t), 0), \quad \mathrm{Ro}(t) = \max(-\dot{s}(t), 0), \\
        \mathrm{Op}(t) &= \max(\dot{o}(t), 0), \quad \mathrm{Cl}(t) = \max(-\dot{o}(t), 0), \\
        \mathrm{Pr}(t) &= \max(\dot{p}(t), 0), \quad \mathrm{Re}(t) = \max(-\dot{p}(t), 0),
    \end{aligned}
\end{equation}
where the subscripts denote Spreading, Rounding, Opening, Closing, Protrusion, and Retraction.

For each sequence $s \in \mathcal{S}$ with frame set $\mathcal{F}_s$, each tendency $k \in \{\mathrm{Sp, Ro, Op, Cl, Pr, Re}\}$ contributes its per-frame values $k(t)$ for $t \in \mathcal{F}_s$. Pooling across all evaluation sequences $\mathcal{S}$, the strength of tendency $k$ is:
\begin{equation}
    S_k = \frac{1}{\sum_{s \in \mathcal{S}} |\mathcal{F}_s|} \sum_{s \in \mathcal{S}} \sum_{t \in \mathcal{F}_s} k(t).
\end{equation}

For each tendency $k$, the model and ground-truth strengths $\hat{S}_k$ and $S_k$ are computed independently from the predicted and ground-truth motion sequences. We then compute their ratio:
\begin{equation}
    r_k = \frac{\hat{S}_k}{S_k},
\end{equation}
and the symmetric GT-normalized similarity:
\begin{equation}
    \mathrm{sim}_k = \exp(-|\log r_k|),
\end{equation}
which equals 1 at exact match ($r_k = 1$) and decays symmetrically for over- ($r_k > 1$) or under-estimation ($r_k < 1$) on a log scale.
The radar in Fig.~\ref{fig3:radar_plot} plots $\mathrm{sim}_k$ for each tendency (Spreading, Rounding, Opening, Closing, Protrusion, Retraction).

\paragraph{\textbf{Quantitative Metrics -- Scales and Units.}}
For clarity, we report the scales and units of all quantitative metrics used in the main paper and supplementary material.
Table~\ref{stab:metric_scales} summarizes the scale factors for each metric across datasets.

\begin{table}[h]
\centering
\small
\setlength{\tabcolsep}{5pt}
\renewcommand{\arraystretch}{1.2}
\caption{Scale factors and units for all reported metrics.}
\vspace{5pt}
\label{stab:metric_scales}
\begin{tabular}{llcc}
\toprule
Metric & Unit & VOCASET & TFHP \\
\midrule
FVE & m & $\times 10^{-3}$ & $\times 10^{-2}$ \\
LVE & m$^2$ & $\times 10^{-4}$ & $\times 10^{-3}$ \\
FDD & m$^2$ & $\times 10^{-8}$ & $\times 10^{-8}$ \\
LDTW & -- & $\times 10^{-3}$ & $\times 10^{-3}$ \\
\midrule
ICW / ILD / LP & m & \multicolumn{2}{c}{$\times 10^{-3}$} \\
ICW / ILD / LP velocity & m/frame & \multicolumn{2}{c}{$\times 10^{-3}$} \\
\bottomrule
\end{tabular}
\end{table}

\paragraph{Dataset Preprocessing.}
VOCASET~\cite{VOCA2019} is originally recorded at 60~fps. We downsample the mesh sequences to 30~fps by selecting every other frame, following the standard preprocessing protocol of prior work~\cite{faceformer2022, codetalker2023, selftalk23}. This ensures that all methods in Table~1 are evaluated under identical temporal resolution. 
Audio is resampled to 16~kHz and converted to mel-spectrograms with a hop length aligned to the 30~fps mesh frame rate.
TFHP~\cite{diffposetalk2024} provides two variants: with head pose and without head pose (FLAME parameters). Since our framework targets articulatory motion and does not model rigid head movement, we use the no-head-pose variant to align the dataset with our method's modeling scope. The dataset distributes per-frame FLAME~\cite{FLAME2017} expression and jaw-pose parameters together with per-speaker shape parameters; we decode these into 5{,}023-vertex meshes using the official FLAME model, with global head rotation and translation set to zero.

\paragraph{\textbf{Hyperparameters}}
We set $\lambda_1 = 10^{-6}$ for $\mathcal{L}_{\mathrm{store}}$ and $\lambda_2 = 10^{3}$ for $\mathcal{L}_{\mathrm{lap}}$ to balance their raw magnitudes against the per-vertex reconstruction loss. $\mathcal{L}_{\mathrm{store}}$, computed in the SAM feature space, is naturally several orders larger than the reconstruction loss, while $\mathcal{L}_{\mathrm{lap}}$, derived from the normalized graph Laplacian, is several orders smaller. The chosen scales bring all loss terms into a comparable range during optimization.

\begin{figure}[H]
\centering
\vspace{-5pt}
\includegraphics[width=0.7\linewidth]{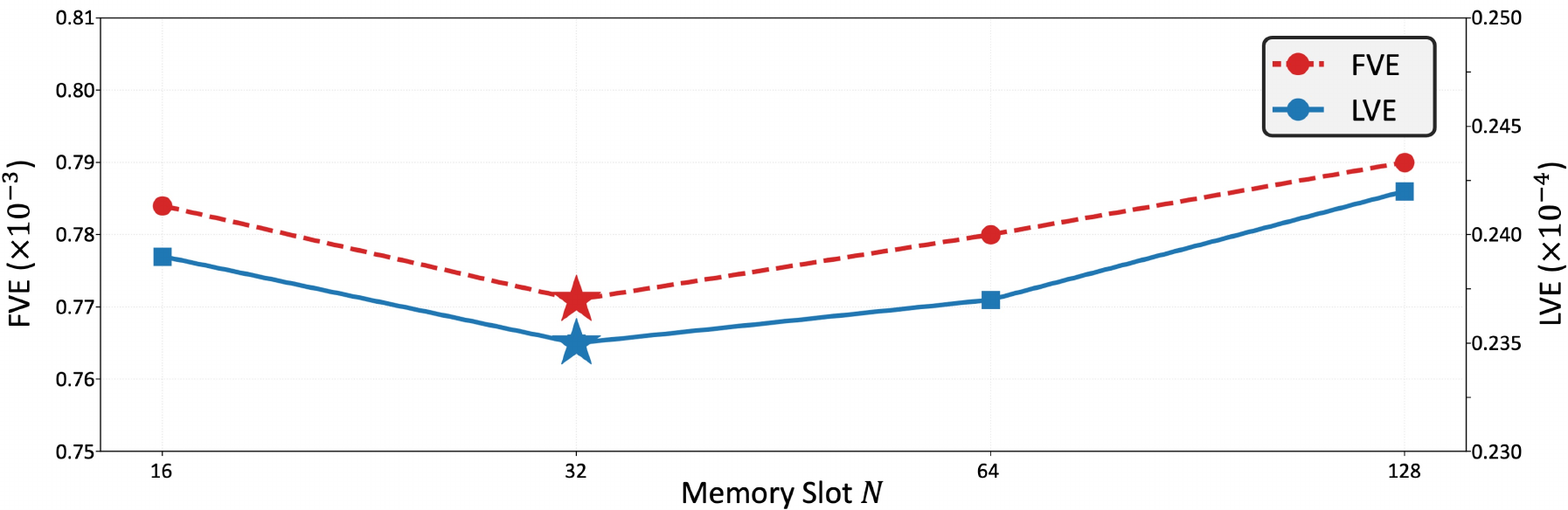}
\caption{FVE and LVE on VOCASET as the number of memory slot $N$. $N=32$ ($\star$) achieves the lowest error on both metrics.}
\label{sfig:sweep}
\end{figure}

We sweep $N \in \{16, 32, 64, 128\}$ on VOCASET, with all other hyperparameters fixed. Fig.~\ref{sfig:sweep} shows that $N=32$ achieves the lowest FVE and LVE, and we adopt this value in all experiments.

\subsection{Ablation Studies}
\label{bc6}

\subsubsection{SAM Ablations}

\begin{figure}[H]
\centering
\includegraphics[width=0.6\linewidth]{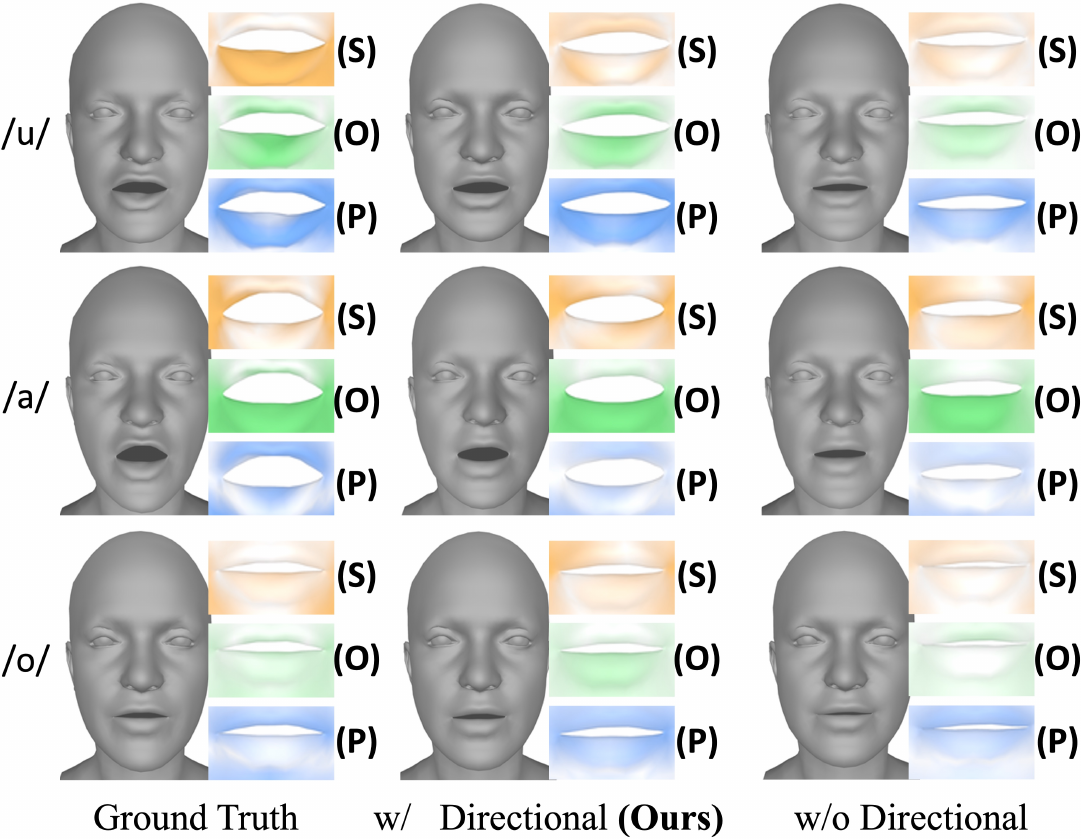}
\caption{Per-phoneme rendered meshes and $S$/$O$/$P$ magnitude heatmaps with and without directional decomposition. Each row shows a representative phoneme (/u/, /a/, /o/).}
\label{fig6:Decomposition_effect}
\end{figure}

\paragraph{\textbf{Effect of Directional Articulatory Decomposition.}}
We replace the three directional branches in SAM with a single unified branch that directly predicts the full facial motion $\mathbf{v}^t \in \mathbb{R}^{V\times 3}$, keeping all other settings unchanged.
Fig.~\ref{fig6:Decomposition_effect} compares both settings on representative phonemes.
With directional decomposition, each heatmap isolates its intended motion and the rendered mesh reflects phoneme-specific lip shapes.
The unified model fails to capture each phoneme's dominant motion: the protrusion of /u/, the opening of /a/, and both the protrusion and spreading of /o/.

As shown in Table~\ref{stab:directional_ablation}, the directional articulatory decomposition reduces FVE by 8.4\%, LVE by 11.7\%, FDD by 20.5\%, and LDTW by 27.1\% over the unified-branch baseline.
This consistent improvement across all metrics demonstrates that decomposing facial motion into three directional articulatory motions ($S$/$O$/$P$) reduces cross-direction interference and improves overall facial dynamics.
\begin{table}[h]
\centering
\small
\setlength{\tabcolsep}{5pt}
\renewcommand{\arraystretch}{1.2}
\vspace{-10pt}
\caption{Ablation on directional articulatory decomposition (VOCASET).}
\vspace{5pt}
\label{stab:directional_ablation}
\begin{tabular}{lcccc}
\toprule
Configuration                   & FVE$\downarrow$   & LVE$\downarrow$   & FDD $\downarrow$  & LDTW $\downarrow$ \\
\midrule
w/o directional articulatory decomposition   & 0.842             & 0.266             & 0.122             & 0.181 \\
\rowcolor{gray!20}
w/ directional articulatory decomposition ($S$/$O$/$P$)    & \textbf{0.771}    & \textbf{0.235}    & \textbf{0.097}    & \textbf{0.132} \\
\bottomrule
\end{tabular}
\end{table}

\paragraph{\textbf{Effect of Memory Structure.}}
A key design of SAM is to store directional articulatory motion priors in explicit learnable memory banks: an acoustic bridge memory $\mathbf{M}_{\mathrm{key}}$ and three per-direction value memories $\mathbf{M}_{\mathrm{val}, S}$, $\mathbf{M}_{\mathrm{val}, O}$, $\mathbf{M}_{\mathrm{val}, P}$.
To verify the necessity of this explicit storage, we remove $\mathbf{M}_{\mathrm{key}}$ and $\mathbf{M}_{\mathrm{val}, \delta}$ while keeping the directional decomposition and TAC unchanged.
Without these memory banks, no retrieval takes place; the acoustic query $\mathbf{q}^t$ and the content feature $\mathbf{f}_{\mathrm{cont}}^t$ are fed directly to each directional articulatory motion decoder $g_\delta(\cdot, \cdot)$ to produce $\hat{\mathbf{v}}_\delta^t$ for $\delta \in \{S, O, P\}$.
As shown in Table~\ref{stab:memory_ablation}, the explicit memory reduces FVE by 6.3\%, LVE by 14.5\%, FDD by 19.8\%, and LDTW by 29.8\% over the no-memory baseline.
Without $\mathbf{M}_{\mathrm{key}}$ and $\mathbf{M}_{\mathrm{val}, \delta}$, the acoustic query and content feature must perform all the disambiguation, and the one-to-many acoustic-to-motion mapping is no longer resolved through stored priors.

\begin{table}[h]
\centering
\small
\setlength{\tabcolsep}{5pt}
\renewcommand{\arraystretch}{1.2}
\vspace{-10pt}
\caption{Ablation on the explicit motion memory in SAM (VOCASET).}
\vspace{5pt}
\label{stab:memory_ablation}
\begin{tabular}{lcccc}
\toprule
Configuration & FVE$\downarrow$ & LVE$\downarrow$ & FDD$\downarrow$ & LDTW$\downarrow$ \\
\midrule
w/o $\mathbf{M}_{\mathrm{key}}, \mathbf{M}_{\mathrm{val}, \delta}$   & 0.823 & 0.275 & 0.121 & 0.188 \\
\rowcolor{gray!20}
Full (Ours)   & \textbf{0.771} & \textbf{0.235} & \textbf{0.097} & \textbf{0.132}\\
\bottomrule
\end{tabular}
\end{table}

\paragraph{\textbf{Effect of Acoustic Query and Content Refinement.}}
SAM uses two complementary speech representations along a directional retrieval pipeline:
an acoustic query from the Mel-spectrogram encoder addresses the per-direction value memory, and a content feature from HuBERT refines the retrieved motion via cross-attention.
To justify this design, we compare three Query--Refinement configurations while keeping the per-direction memory, directional decomposition, and TAC fixed.
Replacing the acoustic query with a content query tests whether HuBERT features alone can index the articulatory memory.
Replacing the content refinement with an acoustic refinement tests whether the Mel features alone can disambiguate the retrieved motion.
As shown in Table~\ref{stab:query_refine_ablation}, both substitutions degrade all metrics, with the largest drop in the Acoustic--Acoustic configuration where content refinement is removed.
Acoustic features address phonetic-context-dependent memory slots more effectively than content features, and content features better disambiguate the retrieved motions.

\begin{table}[h]
\centering
\small
\setlength{\tabcolsep}{5pt}
\renewcommand{\arraystretch}{1.2}
\vspace{-10pt}
\caption{Ablation on the Query--Refinement design in SAM (VOCASET). 
``Acoustic'' denotes Mel-spectrogram features and ``Content'' denotes HuBERT features. 
}
\vspace{5pt}
\label{stab:query_refine_ablation}
\begin{tabular}{llcccc}
\toprule
Query & Refinement & FVE$\downarrow$ & LVE$\downarrow$ & FDD$\downarrow$ & LDTW$\downarrow$ \\
\midrule
Content  & Content              & 0.815 & 0.277 & 0.127 & 0.156 \\
Acoustic & Acoustic             & 0.850 & 0.372 & 0.158 & 0.199 \\
\rowcolor{gray!20}
Acoustic & Content (Ours)       & \textbf{0.771} & \textbf{0.235} & \textbf{0.097} & \textbf{0.132} \\
\bottomrule
\end{tabular}
\end{table}

\paragraph{\textbf{Effect of Value Memory store loss.}}
The store loss $\mathcal{L}_{\mathrm{store}, \delta}$ in Eq.~(\ref{main:store_loss}) aligns each value memory slot with its corresponding ground-truth $(2\Omega+1)$-frame articulatory feature along direction $\delta$.
To assess whether this explicit supervision is necessary, we set $\mathcal{L}_{\mathrm{store}, \delta} = 0$ and retain only the vertex-domain supervisions $\mathcal{L}_{\mathrm{rec}, \delta}$ and $\mathcal{L}_{\mathrm{vel}, \delta}$ from Eq.~(\ref{main:vertex_domain_supervisions}).
Without $\mathcal{L}_{\mathrm{store}}$, the memory slots are shaped only indirectly through retrieval, so nothing prevents different slots from collapsing onto similar motion patterns.
Table~\ref{stab:storing_loss_ablation} shows consistent degradation across all metrics when $\mathcal{L}_{\mathrm{store}}$ is removed.
Without direct slot-level supervision, the value memory no longer organizes into direction-specific motion prototypes.

\begin{table}[h]
\centering
\small
\setlength{\tabcolsep}{5pt}
\renewcommand{\arraystretch}{1.2}
\vspace{-10pt}
\caption{Ablation on the value memory store loss (VOCASET).}
\vspace{5pt}
\label{stab:storing_loss_ablation}
\begin{tabular}{lcccc}
\toprule
Configuration & FVE$\downarrow$ & LVE$\downarrow$ & FDD$\downarrow$ & LDTW$\downarrow$ \\
\midrule
w/o $\mathcal{L}_{\mathrm{store}}$   & 0.811 & 0.267 & 0.117 & 0.144 \\
\rowcolor{gray!20}
w/ $\mathcal{L}_{\mathrm{store}}$ (Ours) & \textbf{0.771} & \textbf{0.235} & \textbf{0.097} & \textbf{0.132}\\
\bottomrule
\end{tabular}
\end{table}

\subsubsection{TAC Ablations}

\paragraph{Topology-aware Composition Effect.}
\label{supp:tac_ablation}
We ablate TAC by replacing it with a topology-independent MLP-based composition (Table~\ref{table:tac_supp_ablation}).
The MLP-based composition processes each vertex in isolation, failing to model spatial coordination across mesh-connected vertices.
In contrast, our GCN-based TAC couples motions across mesh-connected vertices via the facial topology and consistently outperforms the MLP composition on all metrics (FVE: 0.799 $\rightarrow$ 0.771, LVE: 0.243 $\rightarrow$ 0.235, FDD: 0.103 $\rightarrow$ 0.097, LDTW: 0.146 $\rightarrow$ 0.132).
This confirms that topology-aware composition is necessary for translating SAM's per-direction articulatory motions into surface-consistent 3D facial motion.

\begin{table}[h]
\centering
\setlength{\tabcolsep}{5pt}
\renewcommand{\arraystretch}{1.2}
\caption{Topology-independent (MLP) vs.\ topology-aware (Ours) composition on VOCASET.}
\vspace{5pt}
\label{table:tac_supp_ablation}
\begin{tabular}{lcccc}
\toprule
Configuration & FVE$\downarrow$ & LVE$\downarrow$ & FDD$\downarrow$ & LDTW$\downarrow$ \\
\midrule
SAM + MLP composition & 0.799 & 0.243 & 0.103 & 0.146 \\
\rowcolor{gray!20}
SAM + TAC (Ours)      & \textbf{0.771} & \textbf{0.235} & \textbf{0.097} & \textbf{0.132} \\
\bottomrule
\end{tabular}
\end{table}

\paragraph{\textbf{Effect of Internal Residual in TAC.}}
TAC stacks two GCN+TCN blocks with a residual connection between them, letting the final output add the cumulative refinement to the initial projected motion.
We remove this internal residual and process the stack in a purely feedforward manner.
As shown in Table~\ref{stab:residual_ablation}, removing the residual connection degrades all metrics. 
This is consistent with observations in deep graph and temporal networks that residual paths are essential for gradient flow and feature reuse when stacking multiple spatio-temporal operators.

\begin{table}[h]
\centering
\setlength{\tabcolsep}{5pt}
\renewcommand{\arraystretch}{1.2}
\caption{Ablation on the internal residual connection in TAC (VOCASET). The residual path is required for stable optimization across stacked GCN+TCN blocks.}
\vspace{5pt}
\label{stab:residual_ablation}
\begin{tabular}{lcccc}
\toprule
Configuration & FVE$\downarrow$ & LVE$\downarrow$ & FDD$\downarrow$ &LDTW$\downarrow$ \\
\midrule
w/o residual (feedforward)   & 0.782 & 0.238 & 0.111 & 0.134 \\
\rowcolor{gray!20}
w/ residual (Ours)            & \textbf{0.771} & \textbf{0.235} & \textbf{0.097} & \textbf{0.132}\\
\bottomrule
\end{tabular}
\end{table}

\paragraph{\textbf{Effect of Graph Laplacian Consistency Loss.}}
In addition to the reconstruction loss $\mathcal{L}_{\text{rec}}$ and velocity loss $\mathcal{L}_{\text{vel}}$ on the composed motion, we supervise TAC with a graph Laplacian consistency loss $\mathcal{L}_{\text{lap}}$ that encourages the predicted motion to match the ground-truth in terms of local surface curvature.
The Laplacian response $\mathbf{L}\tilde{\mathbf{v}}^t$ measures how each vertex deviates from the normalized mean of its topological neighbors, which corresponds to the local curvature of the facial surface at frame $t$.
Unlike the vertex-wise reconstruction loss, which penalizes only the absolute position error, $\mathcal{L}_{\text{lap}}$ penalizes high-frequency geometric discrepancies, such as mismatched lip-corner folds and creases around the mouth, that are visually salient but carry little weight in global vertex error.
To isolate its effect, we remove $\mathcal{L}_{\text{lap}}$ while keeping the $\mathcal{L}_{\text{rec}}$ and $\mathcal{L}_{\text{vel}}$.
Table~\ref{stab:laplacian_ablation} reports consistent gains across all metrics when $\mathcal{L}_{\text{lap}}$ is added.

\begin{table}[h]
\centering
\setlength{\tabcolsep}{5pt}
\renewcommand{\arraystretch}{1.2}
\caption{Ablation on the graph Laplacian consistency loss (VOCASET). $\mathcal{L}_{\text{lap}}$ constrains local surface curvature and improves overall motion fidelity.}
\vspace{5pt}
\label{stab:laplacian_ablation}
\begin{tabular}{lccccc}
\toprule
Configuration & FVE$\downarrow$ & LVE$\downarrow$ & FDD$\downarrow$ & LDTW$\downarrow$ \\
\midrule
w/o $\mathcal{L}_{\text{lap}}$   & 0.788 & 0.236 & 0.107 & 0.139\\
\rowcolor{gray!20}
w/ $\mathcal{L}_{\text{lap}}$ (Ours) & \textbf{0.771} & \textbf{0.235} & \textbf{0.097} & \textbf{0.132}\\
\bottomrule
\end{tabular}
\end{table}

\subsubsection{Training Strategy}
\paragraph{\textbf{Effect of Two-stage Training Strategy.}}
Our training schedule first pretrains SAM in Stage~1 for 250 epochs and then freezes SAM while training TAC in Stage~2 for 150 epochs.
To validate this choice, we compare two alternatives.
In \textit{unfrozen fine-tuning}, Stage~1 is loaded from the same pretrained checkpoint but left unfrozen, so SAM and TAC are optimized jointly with the Stage~2 losses.
In \textit{unified single-stage training}, Stage~1 is not pretrained at all; SAM and TAC are trained jointly from scratch with the full objective.
Table~\ref{stab:training_strategy_ablation} shows that both alternatives underperform our frozen two-stage schedule.
Unified single-stage training suffers the largest drop: the directional motion prototypes in SAM require dedicated supervision before being consumed by TAC.
Unfrozen fine-tuning is closer to ours but still degrades, indicating that keeping SAM frozen after convergence preserves its direction-specific retrieval.

\begin{table}[h]
\centering
\small
\setlength{\tabcolsep}{5pt}
\renewcommand{\arraystretch}{1.2}
\caption{Ablation on the training schedule (VOCASET). Pretraining SAM and freezing it during TAC training outperforms both end-to-end fine-tuning and unified single-stage training.}
\vspace{5pt}
\label{stab:training_strategy_ablation}
\begin{tabular}{lcccc}
\toprule
Configuration & FVE$\downarrow$ & LVE$\downarrow$ & FDD$\downarrow$ & LDTW$\downarrow$\\
\midrule
Unified single-stage (from scratch)    & 0.801 & 0.245 & 0.118 & 0.142 \\
Pretrained + unfrozen fine-tuning      & 0.798 & 0.244 & 0.122 & 0.144 \\
\rowcolor{gray!20}
Pretrained + frozen (Ours)              & \textbf{0.771} & \textbf{0.235} & \textbf{0.097} & \textbf{0.132}\\
\bottomrule
\end{tabular}
\end{table}

\subsubsection{Hyperparameter Sensitivity}
\label{bc7}

\begin{figure}[H]
\centering
\includegraphics[width=0.8\linewidth]{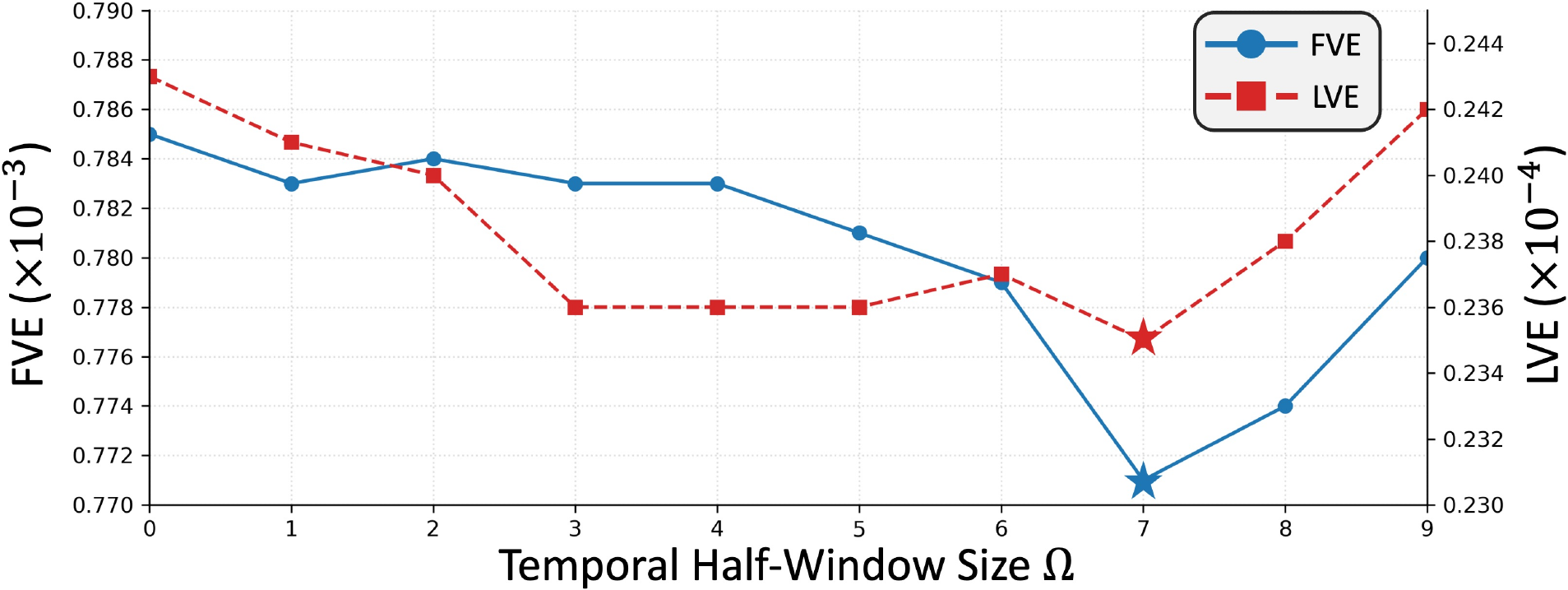}
\caption{Effect of half-window size $\Omega$ on FVE and LVE (VOCASET). $\Omega{=}0$ corresponds to frame-independent retrieval without temporal motion context.}
\label{sfig:omega}
\end{figure}

\paragraph{\textbf{Effect of Temporal Window Size $\Omega$.}}
We investigate the effect of the half-window size $\Omega$ in the value memory on reconstruction performance, where $\Omega$ defines the temporal context span of $2\Omega+1$ frames.
Fig.~\ref{sfig:omega} shows FVE and LVE on VOCASET as $\Omega$ varies over $\{0, 1, 2, 3, 4, 5, 6, 7, 8, 9\}$.
Our method achieves the lowest FVE and LVE at $\Omega{=}7$, where the $2\Omega+1$-frame window provides sufficient surrounding phonetic context for stable and disambiguated retrieval.
This sweet spot consistently outperforms both shorter windows ($\Omega{\leq}1$), which lack temporal motion context, and longer windows ($\Omega{\geq}9$), which introduce temporally distant information less relevant to the current articulatory event.
Based on this analysis, we set $\Omega{=}7$ in all experiments.

\subsection{Generalization Studies}

\paragraph{\textbf{Topology Generalization on BIWI.}}
To validate that our method generalizes beyond FLAME topology, we evaluate on BIWI~\cite{BIWI2010}, whose mesh topology (23{,}370 vertices) differs substantially from FLAME (5{,}023 vertices).
BIWI comprises 14 speakers each uttering 40 sentences twice (neutral and emotional) at 25~fps, with an average sequence length of 4.67~seconds. Following prior work~\cite{faceformer2022, codetalker2023}, we adopt the BIWI-Train (192 sentences, 6 subjects), -Val (24 sentences, 6 subjects), -Test-A (24 sentences, 6 subjects), and -Test-B (32 sentences, 8 subjects) splits. We train SAM+TAC with the same optimizer, learning rate, batch size, and training schedule as on VOCASET and TFHP, setting $N=64$, $D_{\mathrm{SAM}}=256$, and $D_{\mathrm{TAC}}=64$. As shown in Table~\ref{stab:biwi}, our full model achieves the best FVE, LVE, and LDTW on BIWI.

\begin{table}[h]
\centering
\small
\setlength{\tabcolsep}{4pt}
\renewcommand{\arraystretch}{1.2}
\caption{Topology generalization on BIWI (23,370 vertices).}
\label{stab:biwi}
\begin{tabular}{lcccc}
\toprule
Methods & FVE ($\times 10^{-1}$)$\downarrow$ & LVE ($\times 10^{-2}$)$\downarrow$ & FDD ($\times 10^{-5}$)$\downarrow$ & LDTW ($\times 10^{-2}$)$\downarrow$ \\
\midrule
CodeTalker   & 0.124 & 0.127 & 0.461 & 0.319 \\
SelfTalk     & 0.120 & 0.114 & \textbf{0.148} & 0.285 \\
UniTalker    & \underline{0.112} & \textbf{0.106} & 0.353 & \underline{0.283} \\
Ours (SAM only)   & 0.115 & 0.113 & 0.557 & 0.296 \\
\rowcolor{gray!20}
Ours (SAM+TAC)    & \textbf{0.110} & \textbf{0.106} & \underline{0.351} & \textbf{0.275} \\
\bottomrule
\end{tabular}
\end{table}

\paragraph{\textbf{Articulatory Errors on TFHP.}}
\label{sec:articulatory_error_tfhp}
We additionally report visible articulatory distance and velocity errors on TFHP in Table~\ref{stab:tfhp_articulatory}. Our method achieves the lowest errors on all three signals (ICW, ILD, LP) for both distance and velocity. These results generalize our VOCASET findings to TFHP, which spans various speaking styles.

\begin{table}[h]
\centering
\small
\setlength{\tabcolsep}{4pt}
\renewcommand{\arraystretch}{1.2}
\caption{Visible articulatory distance and velocity errors on TFHP.}
\label{stab:tfhp_articulatory}
\begin{tabular}{lccc|ccc}
\toprule
\multirow{2}{*}{Methods}
& \multicolumn{3}{c|}{Distance Error}
& \multicolumn{3}{c}{Velocity Error} \\
& ICW$\downarrow$ & ILD$\downarrow$ & LP$\downarrow$
& ICW$\downarrow$ & ILD$\downarrow$ & LP$\downarrow$ \\
\midrule
Mimic  & \underline{4.320} & \underline{5.326} & \underline{5.968} & {1.460} & 2.593 & {2.871} \\
DiffPoseTalk  & {5.483} & {5.860} & {7.006} & \underline{1.221} & \underline{2.168} & \underline{2.438} \\
ARTalk        & 5.788 & 6.049 & 7.378 & {1.501} & 2.801 & {2.644} \\
\rowcolor{gray!20}
\textbf{Ours} & \textbf{4.299} & \textbf{4.829} & \textbf{5.859} & \textbf{1.168} & \textbf{2.083} & \textbf{2.414} \\
\bottomrule
\end{tabular}
\end{table}

\subsection{Qualitative Analysis}

\begin{figure}[H]
\centering
\includegraphics[width=1.0\linewidth]{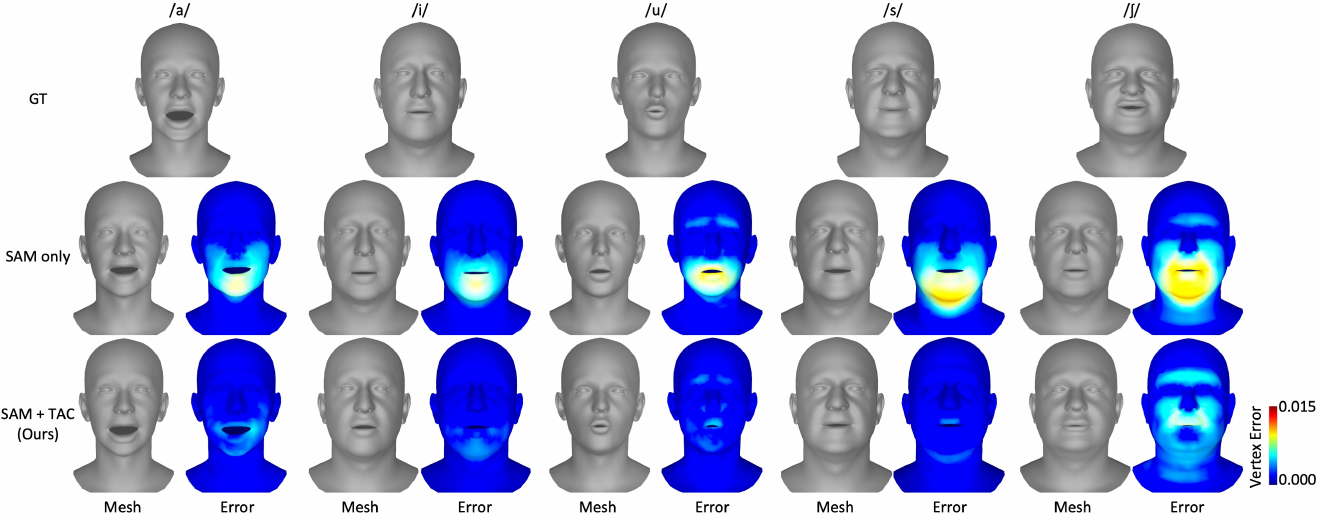}
\caption{Stage-wise qualitative comparison on TFHP. Top row: GT. Middle row: Stage 1 (SAM only) result and per-vertex error. Bottom row: Stage 2 (SAM + TAC) result and per-vertex error.}
\label{sfig:tac_stage_comparison}
\end{figure}

\paragraph{\textbf{Stage-wise Effect of TAC.}}
Fig.~\ref{sfig:tac_stage_comparison} presents a stage-wise qualitative comparison on TFHP, complementing the quantitative composition ablation in Table~\ref{table:module_ablation_revised} of the main paper with per-vertex error visualization.
For five representative phonemes, we compare Stage 1 (SAM only) and Stage 2 (SAM + TAC) against the ground truth, color-coded by per-vertex L1 error.
SAM only already produces phoneme-consistent lip shapes through directional articulatory motions, but residual errors remain concentrated around the lower face (lips, chin, jaw line).
Adding TAC in Stage 2 reduces these residuals by composing the directional articulatory motions over the mesh topology, producing surface-coherent reconstructions that align more tightly with the ground truth.

\begin{figure}[H]
\centering
\includegraphics[width=1.0\linewidth]{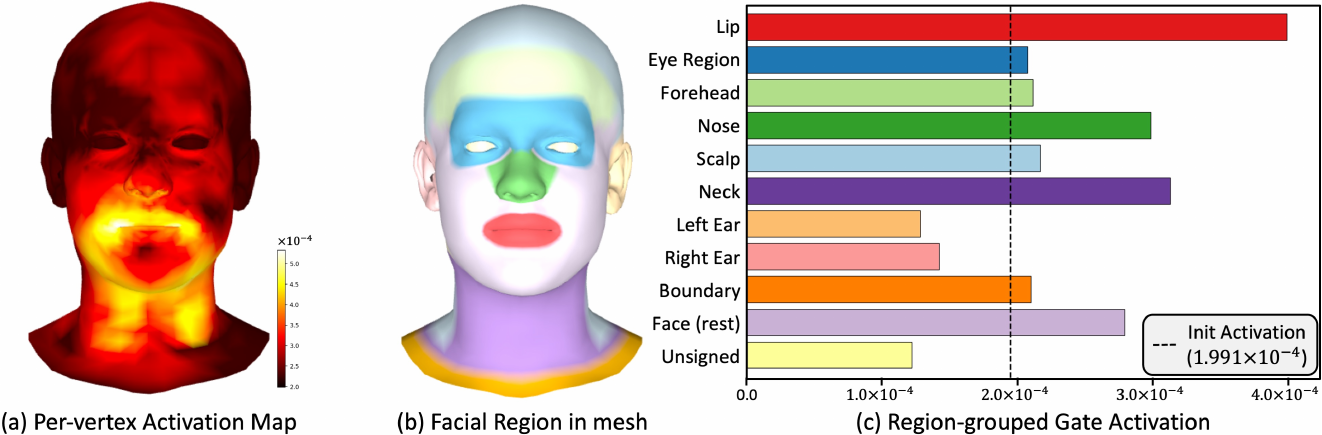}
\caption{Visualization of the softmax distribution $\mathrm{softmax}(\tau\boldsymbol{\alpha})$ underlying the TAC modulation gate $\mathbf{m}$ on the FLAME template. (a) Per-vertex distribution. (b) Anatomical region segmentation. (c) Region-grouped mean, with the dashed line indicating the uniform baseline $1/V$.}
\label{sfig:gate_hotspot}
\end{figure}

\paragraph{\textbf{Vertex-wise Gate Visualization.}}
To verify whether the TAC modulation gate $\mathbf{m} = V\cdot\mathrm{softmax}(\tau\boldsymbol{\alpha})$ forms a meaningful spatial structure, we visualize its underlying softmax distribution $\mathrm{softmax}(\tau\boldsymbol{\alpha})$ on the FLAME template (Fig.~\ref{sfig:gate_hotspot}~(a)) and aggregate it by anatomical region (Fig.~\ref{sfig:gate_hotspot}~(b, c)).
Since $\mathrm{softmax}(\tau\boldsymbol{\alpha})$ sums to 1 across vertices, the uniform baseline is $1/V$, and any region above $1/V$ indicates higher allocation than a uniform distribution.
The Lip region receives the highest allocation, reaching $2.0\times$ the uniform baseline $1/V$ and clearly above all other regions.
Neck, Nose, and the remaining face area also exceed the baseline, consistent with their proximity to articulation-driven deformation, while the forehead, scalp, eye region, and ears stay near or below the baseline as they remain largely static during speech.
This spatially structured allocation emerges without any explicit supervision and aligns with the lip-jaw deformability prior of TAC's design, demonstrating that the gate captures articulation-aware motion structure consistent with the deformability of facial regions.

\begin{figure}[H]
\centering
\includegraphics[width=1.0\linewidth]{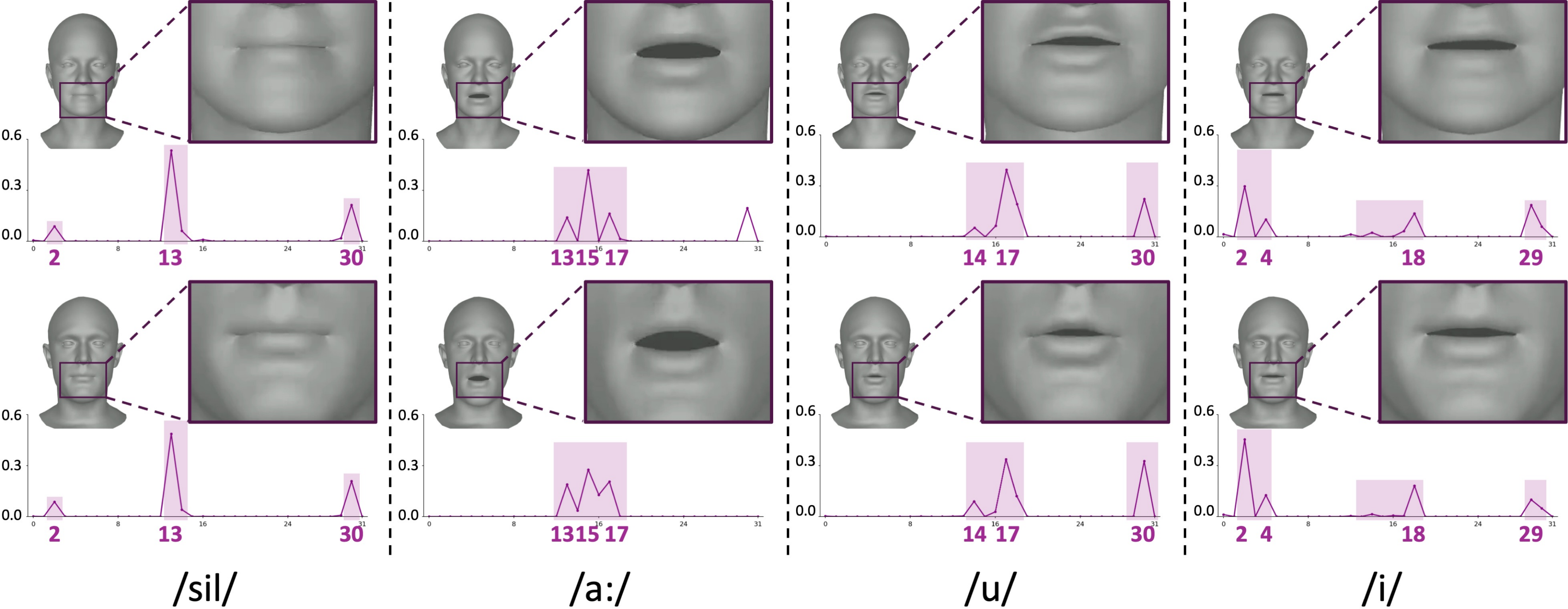}
\caption{Consistency of acoustic-bridge memory addressing vectors. For the same phoneme class, the addressing weights repeatedly peak at similar slots across occurrences, suggesting stable, reusable retrieval.}
\label{sfig:viz_acoustic_mem}
\end{figure}

\paragraph{\textbf{Acoustic-Bridge Memory Addressing.}}
Fig.~\ref{sfig:viz_acoustic_mem} visualizes the addressing vectors of the acoustic-bridge memory.
For the same phoneme class (e.g., /sil/, /a:/, /u/, /i/), the addressing weights peak at similar slots across different occurrences, indicating phoneme-consistent retrieval.

\begin{figure}[H]
\centering
\includegraphics[width=1.0\linewidth]{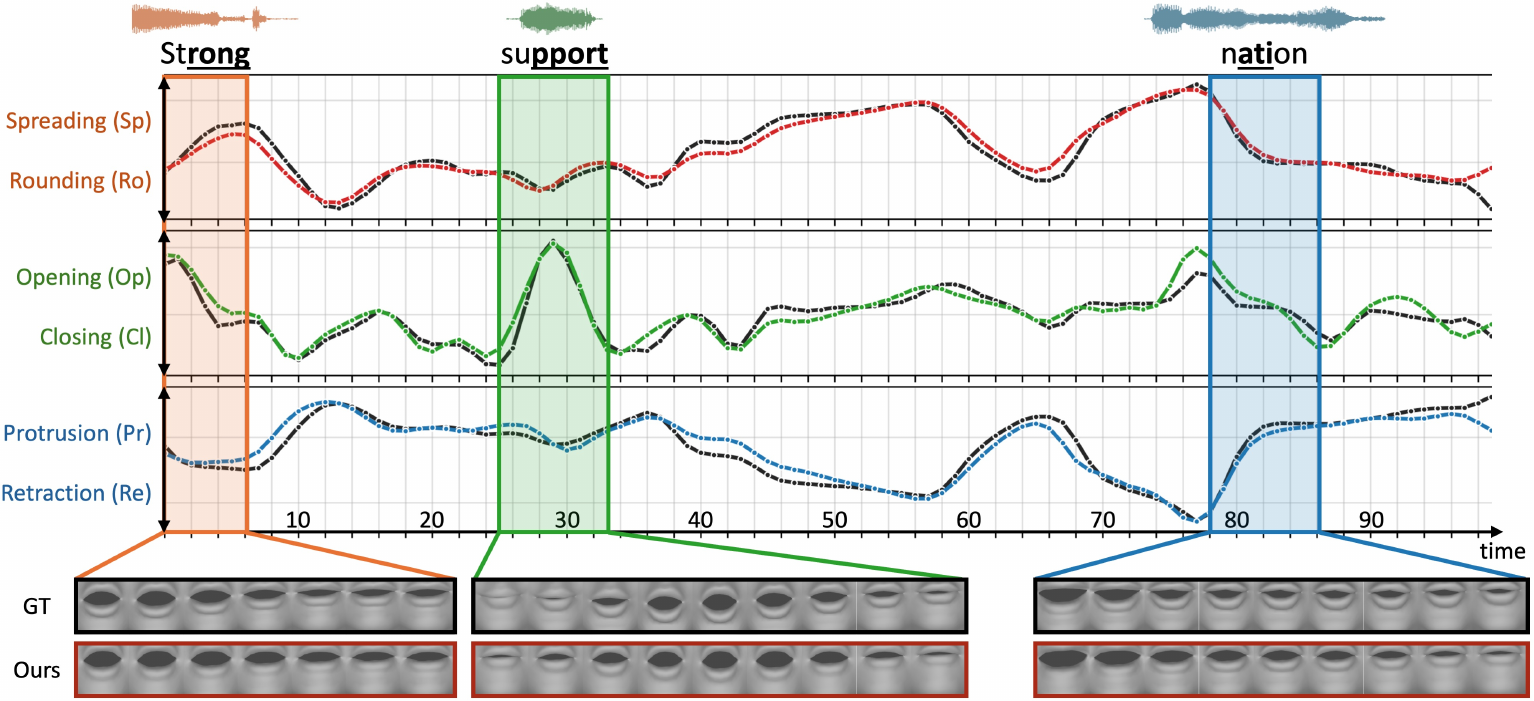}
\caption{Visible articulatory lip-motion trajectories of our method and the ground truth along three directions (Spreading--Rounding, Opening--Closing, Protrusion--Retraction). Highlighted segments correspond to the words ``Strong'', ``support'', and ``nation'' in the utterance, with GT and Ours mesh sequences shown below.}
\label{sfig:trajectory_ours}
\end{figure}

\begin{figure}[H]
\centering
\includegraphics[width=1.0\linewidth]{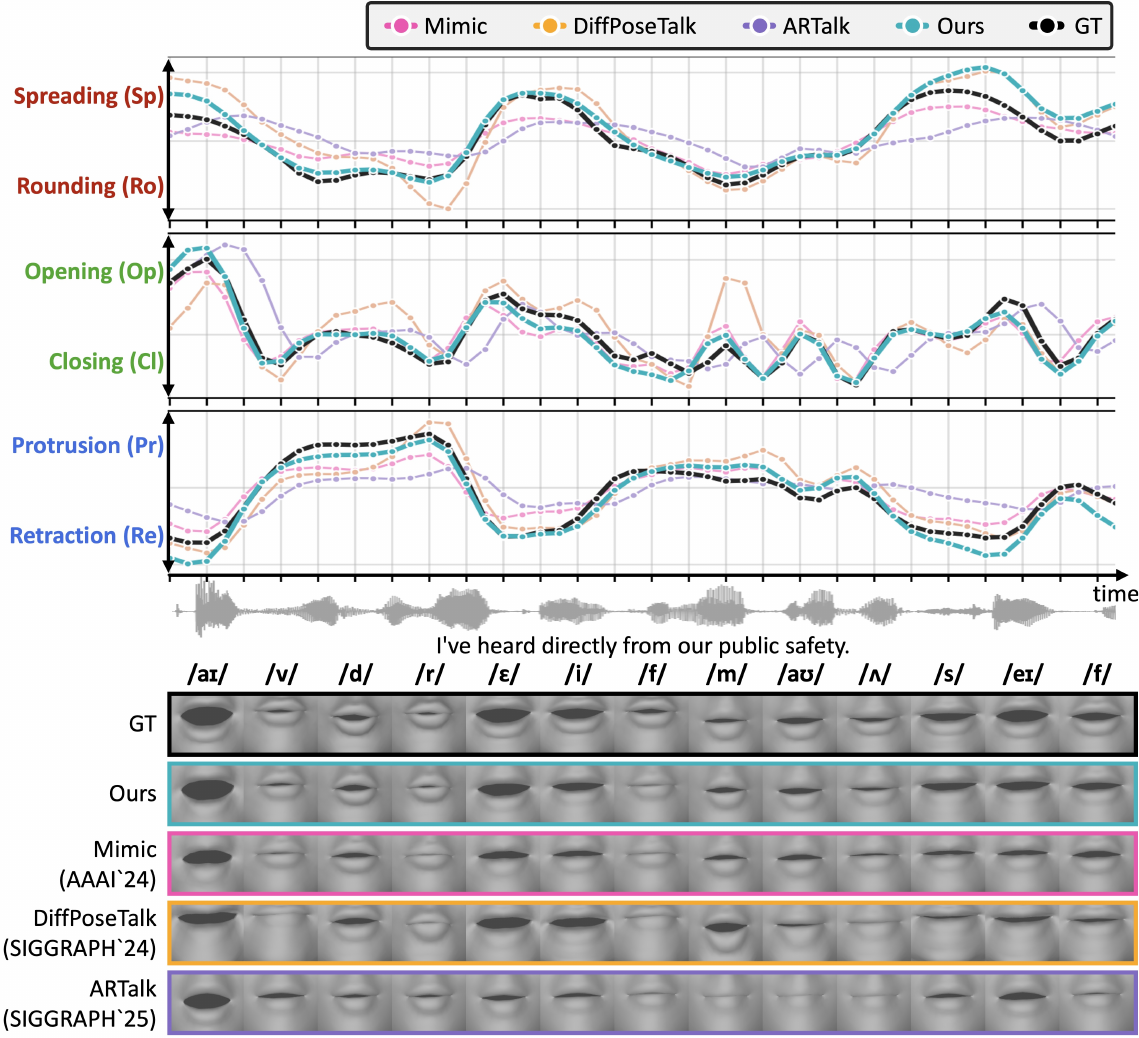}
\caption{Additional example of three directional articulatory trajectories (Spreading--Rounding, Opening--Closing, Protrusion--Retraction) for the utterance ``I've heard directly from our public safety.''}
\label{sfig:trajectory}
\end{figure}

\paragraph{\textbf{Articulatory Motion Trajectory.}}
\label{ssec:Articulatory_Motion_Trajectory}
Fig.~\ref{sfig:trajectory_ours} visualizes our method's three directional motion trajectories alongside the ground truth.
The trajectories are derived from four lip-surface anchors (UL, LL, LMC, RMC) and capture articulatory patterns at three word segments: spreading at ``Strong'', opening followed by closing at ``support'', and combined opening and retraction at ``nation''.
Across these segments, our trajectories follow the ground truth, and the corresponding mesh sequences reproduce phoneme-specific lip shape.
Fig.~\ref{sfig:trajectory} shows an additional example with all baseline methods on a different utterance, where the direction-specific deviations observed in Fig.~\ref{fig6:trajectory_analysis} of the main paper recur.

\begin{figure}[H]
\centering
\includegraphics[width=1.0\linewidth]{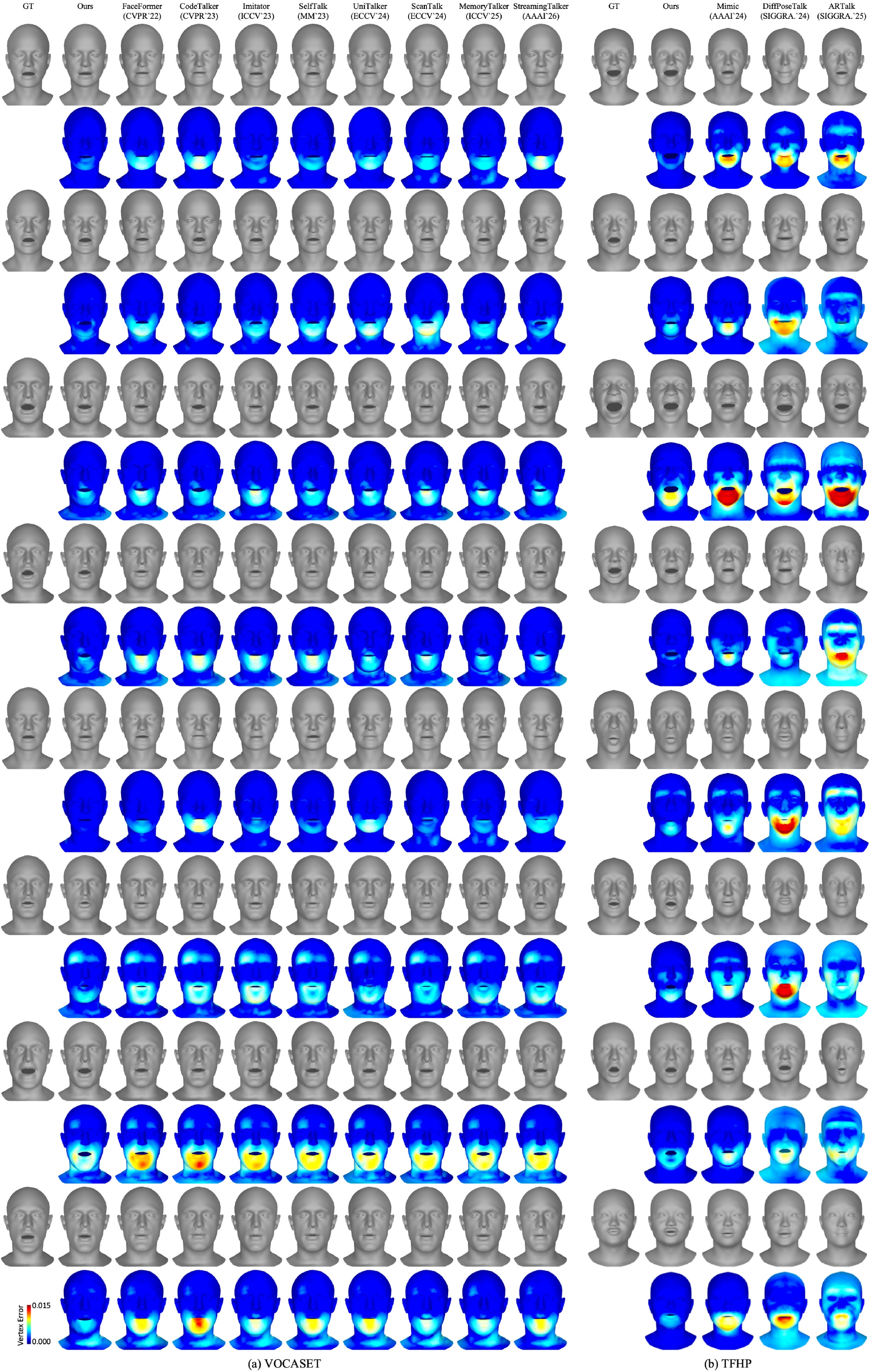}
\caption{Additional qualitative comparisons on VOCASET and TFHP.}
\vspace{-20pt}
\label{sfig:qualitative_eval}
\end{figure}

\paragraph{\textbf{Comparisons with State-of-the-Art.}}
Fig.~\ref{sfig:qualitative_eval} presents additional frame-level visual comparisons on VOCASET and TFHP.
The first and third rows display full-face renderings, while the second and fourth rows show the per-vertex L1 error against the ground truth.
Our method produces distinct articulatory configurations across phonemes, while prior methods tend to underestimate the magnitude of these motions.
Across both datasets, our error maps remain low at the lips and lower-face region (jaw, chin), indicating that articulation-aware modeling improves articulatory reconstruction.
The upper-face region remains comparable to competing methods, indicating that our model captures articulatory motion without affecting non-articulatory regions.

\subsection{User Study Details}
\label{supp:user_study_protocol}
We design our user study following the pairwise A/B preference protocol of~\cite{kim2025memorytalker} for perceptual evaluation of speech-driven 3D facial animation.

\paragraph{Recruitment.}
We recruited 26 voluntary participants through our extended network (e.g., colleagues and acquaintances). Two participants who failed attention checks were excluded, resulting in 24 valid participants.

\paragraph{Compensation.}
Participation was voluntary, and no compensation was provided. This was disclosed to all participants prior to obtaining consent.

\paragraph{Informed Consent.}
The study was administered via a Google Forms questionnaire. Before responding, all participants were presented with a description of (i) the task: viewing short side-by-side video clips of 3D facial animations and selecting preferences, (ii) the expected duration, (iii) the data being recorded (forced-choice responses only), (iv) the absence of compensation, and (v) the right to withdraw at any time. Only participants who acknowledged the description proceeded to the questionnaire.

\paragraph{Privacy and Anonymization.}
We did not collect any personally identifiable information. The Google Forms questionnaire was configured to disable email collection and login requirements, so respondents were anonymous to the study administrators. Only forced-choice responses were recorded; no individual response can be associated with a specific participant's identity.

\begin{figure}[H]
\centering
\includegraphics[width=1.0\linewidth]{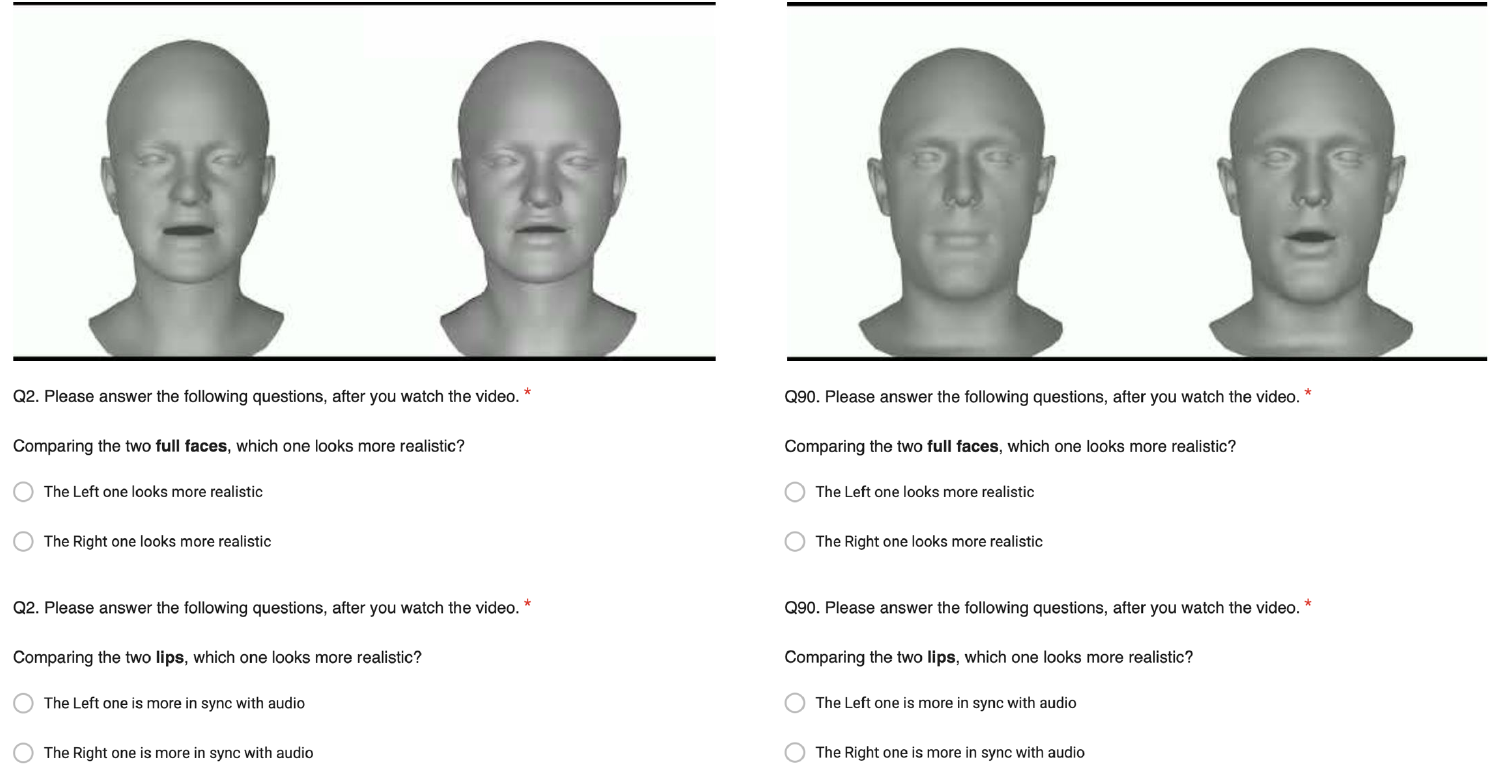}
\caption{Examples of the conducted user study.}
\label{sfig:userstudy}
\end{figure}

\paragraph{Task and Instructions.}
Participants viewed pairs of short video clips of speech-driven 3D facial animations, presented side-by-side in randomized order. For each pair, participants were asked two forced-choice questions: (1) which animation showed better lip-sync accuracy with the audio, and (2) which animation appeared more realistic overall. Each participant rated 15 randomly sampled clips per competitor for each of the seven competitor methods. To identify inattentive participants, we inserted 10 attention-check trials at random positions, in which one video was synchronized with the audio while the other was deliberately desynchronized. Participants who failed to consistently identify the synchronized video were excluded. Representative interface screenshots are shown in Fig.~\ref{sfig:userstudy}.

\paragraph{Ethical Review.}
The study consists of a pairwise preference test where participants only viewed short 3D talking-head video clips. It does not involve the collection of personal or sensitive information and poses minimal risk. IRB approval was not obtained for this study.

\subsection{Limitations and Broader Impacts}
\label{bc8}

\paragraph{\textbf{Limitations.}}
\label{sec:limitations}
Our method has two limitations that suggest directions for future work. First, we use the FLAME mesh topology (5,023 vertices) and BIWI mesh topology (23,370 vertices), neither of which includes intra-oral structures such as teeth and tongue. While the three directional articulatory motions control lip opening and thus influence when these structures become visible, explicitly modeling them would improve realism for phonemes that depend on tongue placement (e.g., /\texttheta/).
Second, our framework generates 3D facial motion from speech alone and does not explicitly model non-articulatory upper-face dynamics such as blinking. 
This limitation is reflected in FDD on TFHP, where blinking is frequent. 
DiffPoseTalk reaches a lower FDD (0.321) by modeling such dynamics through diffusion-based stochastic sampling, while our audio-only method (0.578) and ARTalk (0.581) attain comparable FDD without it. 
Table~\ref{stab:region_fdd} shows that methods without blinking modeling (Ours and Mimic) exhibit nearly identical eye-region FDD, whereas DiffPoseTalk and ARTalk achieve lower values through stochastic and autoregressive blinking generation respectively. 
Non-blinking regions (forehead, scalp) show minimal cross-method variation. 
On VOCASET, recorded under controlled conditions with limited upper-face activity, our method achieves the best FDD (0.097), confirming that accurate articulation does not come at the cost of upper-face degradation. 
Extending our framework with a stochastic blinking module remains future work.

\begin{table}[h]
\centering
\small
\setlength{\tabcolsep}{8pt}
\renewcommand{\arraystretch}{1.15}
\caption{Per-region FDD on TFHP ($\times 10^{-7}$ m$^2$, lower is better). Regions follow the official FLAME topology (Fig.~\ref{sfig:gate_hotspot} (b)).}
\vspace{5pt}
\label{stab:region_fdd}
\begin{tabular}{lccc}
\toprule
\multirow{2}{*}{Method} & Forehead$\downarrow$ & Eyes$\downarrow$ & Scalp$\downarrow$ \\
 & (133 verts) & (574 verts) & (489 verts) \\
\midrule
Mimic         & 0.020 & 0.099 & 0.010 \\
DiffPoseTalk  & \textbf{0.015} & \textbf{0.042} & \textbf{0.008} \\
ARTalk        & 0.044 & 0.060 & 0.011 \\
\rowcolor{gray!20}
Ours          & 0.023 & 0.102 & 0.010 \\
\bottomrule
\end{tabular}
\end{table}

\paragraph{\textbf{Broader Impacts.}}
Our framework advances speech-driven 3D facial animation with potential positive applications in virtual reality, education, accessibility for hearing-impaired individuals, and content production.
However, improved facial animation quality also raises the risk of misuse for generating deceptive content such as deepfakes.
We encourage the development of complementary detection methods and responsible release practices alongside deployment of such technologies.

\newpage
\section*{NeurIPS Paper Checklist}

\begin{enumerate}

\item {\bf Claims}
    \item[] Question: Do the main claims made in the abstract and introduction accurately reflect the paper's contributions and scope?
    \item[] Answer: \answerYes{}
    \item[] Justification: The Abstract and Sec.~\ref{main:intro} explicitly state our contributions and scope. Specifically, we (1) propose a first articulation-aware framework for speech-driven 3D facial animation grounded in structured visible articulation; (2) introduce Speech--Articulatory Memory (SAM) to model speech-to-articulation correspondence via key-value memory retrieval and decoding under phonetic context; (3) introduce Topology-aware Articulatory Composition (TAC) to compose directional articulatory motions into coherent 3D facial motion under mesh topology; and (4) demonstrate the effectiveness of SAM and TAC through quantitative, qualitative, and user-study evaluations on VOCASET and TFHP, achieving state-of-the-art performance on standard reconstruction metrics and improved articulatory fidelity.

\item {\bf Limitations}
    \item[] Question: Does the paper discuss the limitations of the work performed by the authors?
    \item[] Answer: \answerYes{}
    \item[] Justification: See Appendix~\ref{sec:limitations}, which discusses two limitations: (1) the FLAME and BIWI mesh topologies do not include intra-oral structures (teeth, tongue), limiting realism for phonemes that depend on tongue placement; and (2) our formulation targets speech-driven articulatory motion and does not explicitly model upper-face dynamics such as blinking and idle head motion, which leaves a remaining FDD gap on TFHP relative to DiffPoseTalk's stochastic upper-face sampling.

\item {\bf Theory assumptions and proofs}
    \item[] Question: For each theoretical result, does the paper provide the full set of assumptions and a complete (and correct) proof?
    \item[] Answer: \answerNA{}
    \item[] Justification: This is an empirical paper. It does not include theorems, lemmas, or formal proofs; all equations describe the proposed network architecture and training objectives.

    \item {\bf Experimental result reproducibility}
    \item[] Question: Does the paper fully disclose all the information needed to reproduce the main experimental results of the paper to the extent that it affects the main claims and/or conclusions of the paper (regardless of whether the code and data are provided or not)?
    \item[] Answer: \answerYes{}
    \item[] Justification: Sec.~\ref{bc1} fully describes the SAM and TAC architectures. Sec.~\ref{bc2} specifies the datasets and train/val/test splits for VOCASET and TFHP.  Sec.~\ref{bc3} reports all training details (Adam optimizer with learning rate $10^{-4}$, batch size 1, two-stage schedule of 250+150 epochs, memory slots $N=32$, half-window $\Omega=7$, $D_{\mathrm{SAM}}=128$, $D_{\mathrm{TAC}}=64$, single RTX 4090). Sec.~\ref{bc4} defines the evaluation metrics, and Appendix~\ref{supp:Implementation_Details} and \ref{supp:tendency_analysis} provide additional implementation and evaluation details (motion autoencoder architecture, autoencoder training, and periodic positional encoding for SAM).

\item {\bf Open access to data and code}
    \item[] Question: Does the paper provide open access to the data and code, with sufficient instructions to faithfully reproduce the main experimental results, as described in supplemental material?
    \item[] Answer: \answerNo{}
    \item[] Justification: All datasets used in this paper (VOCASET and TFHP for the main experiments, BIWI for supplementary topology-generalization analysis) are publicly available, and we follow the standard preprocessing and split protocols established in prior work~\cite{faceformer2022, codetalker2023, diffposetalk2024}. Sec.~\ref{bc1}, Sec.~\ref{bc3}, and Appendix~\ref{bc5} disclose the full architecture, hyperparameters, and training schedule needed to re-implement the method. Code will be released upon acceptance.

\item {\bf Experimental setting/details}
    \item[] Question: Does the paper specify all the training and test details (e.g., data splits, hyperparameters, how they were chosen, type of optimizer) necessary to understand the results?
    \item[] Answer: \answerYes{}
    \item[] Justification: Sec.~\ref{bc2} specifies dataset splits following standard protocols~\cite{faceformer2022, codetalker2023, diffposetalk2024}. Sec.~\ref{bc3} reports the optimizer (Adam, lr $10^{-4}$, batch size 1), the two-stage training schedule, and all architectural hyperparameters ($N$, $\Omega$, $D_{\mathrm{SAM}}$, $D_{\mathrm{TAC}}$). Sec.~\ref{bc4} defines the four standard reconstruction metrics (FVE, LVE, FDD, LDTW) and the visible articulatory distance and velocity errors used at test time. Appendix Sec.~\ref{bc6} provides ablation studies and a sensitivity analysis (Appendix~\ref{bc7}) that justify the chosen design and hyperparameter values.

\item {\bf Experiment statistical significance}
    \item[] Question: Does the paper report error bars suitably and correctly defined or other appropriate information about the statistical significance of the experiments?
    \item[] Answer: \answerYes{}
    \item[] Justification: For the user study (Sec.~\ref{main:userstudy}), we report a one-sided binomial test against the 50\% chance level on $24$ valid participant responses (after excluding $2$ failed attention checks from $26$ recruits), each rating $15$ randomly sampled clips per competitor; all seven pairwise comparisons yield $p < 0.0001$.

\item {\bf Experiments compute resources}
    \item[] Question: For each experiment, does the paper provide sufficient information on the computer resources (type of compute workers, memory, time of execution) needed to reproduce the experiments?
    \item[] Answer: \answerYes{}
    \item[] Justification: Sec.~\ref{bc3} reports that all experiments are conducted on a single NVIDIA RTX 4090 GPU (24~GB VRAM) with batch size 1, using a two-stage schedule of 250 epochs (Stage~1, SAM) and 150 epochs (Stage~2, TAC).
    
\item {\bf Code of ethics}
    \item[] Question: Does the research conducted in the paper conform, in every respect, with the NeurIPS Code of Ethics \url{https://neurips.cc/public/EthicsGuidelines}?
    \item[] Answer: \answerYes{}
    \item[] Justification: The research uses only publicly available datasets under their release terms, and the user study collects only pairwise preference responses without personally identifiable information. Potential misuse risks (e.g., deepfakes) are discussed in Appendix~\ref{bc8}.

\item {\bf Broader impacts}
    \item[] Question: Does the paper discuss both potential positive societal impacts and negative societal impacts of the work performed?
    \item[] Answer: \answerYes{}
    \item[] Justification: Appendix~\ref{bc8} discusses both positive societal impacts (applications in virtual reality, education, accessibility for hearing-impaired individuals, and content production) and negative impacts (risk of misuse for generating deceptive content such as deepfakes), along with mitigation strategies including the development of complementary detection methods and responsible release practices.
    
\item {\bf Safeguards}
    \item[] Question: Does the paper describe safeguards that have been put in place for responsible release of data or models that have a high risk for misuse (e.g., pre-trained language models, image generators, or scraped datasets)?
    \item[] Answer: \answerNA{}
    \item[] Justification: The proposed model outputs 3D vertex motion sequences rather than photorealistic content, and therefore does not fall into the high-risk category described in this question (e.g., pre-trained language models, image/video generators, or scraped datasets). Broader misuse considerations and recommended responsible release practices are discussed in Appendix~\ref{bc8}.

\item {\bf Licenses for existing assets}
    \item[] Question: Are the creators or original owners of assets (e.g., code, data, models), used in the paper, properly credited and are the license and terms of use explicitly mentioned and properly respected?
    \item[] Answer: \answerYes{}
    \item[] Justification: All datasets (VOCASET~\cite{VOCA2019}, TFHP~\cite{diffposetalk2024}, HDTF~\cite{zhang2021flow}, BIWI~\cite{BIWI2010}), the FLAME mesh template~\cite{FLAME2017}, and the pre-trained HuBERT model~\cite{HuBERT2021} are cited with their original publications and used in accordance with their respective release terms for non-commercial research.

\item {\bf New assets}
    \item[] Question: Are new assets introduced in the paper well documented and is the documentation provided alongside the assets?
    \item[] Answer: \answerNA{}
    \item[] Justification: The paper does not introduce new datasets or benchmarks. The code supporting the proposed method will be released upon acceptance with appropriate documentation.

\item {\bf Crowdsourcing and research with human subjects}
    \item[] Question: For crowdsourcing experiments and research with human subjects, does the paper include the full text of instructions given to participants and screenshots, if applicable, as well as details about compensation (if any)? 
    \item[] Answer: \answerYes{}
    \item[] Justification: Sec.~\ref{main:userstudy} reports the user study protocol (26 participants recruited, 24 valid responses after attention checks, 15 randomly sampled clips per competitor, pairwise side-by-side A/B preference task). Appendix~\ref{supp:user_study_protocol} provides representative screenshots of the questionnaire interface. Participation was voluntary, with no monetary compensation provided.

\item {\bf Institutional review board (IRB) approvals or equivalent for research with human subjects}
    \item[] Question: Does the paper describe potential risks incurred by study participants, whether such risks were disclosed to the subjects, and whether Institutional Review Board (IRB) approvals (or an equivalent approval/review based on the requirements of your country or institution) were obtained?
    \item[] Answer: \answerNo{}
    \item[] Justification: The study consists of a simple pairwise preference test. The participants only viewed short 3D talking-head video clips. It does not involve the collection of personal or sensitive information and poses minimal risk. IRB approval was not required.

\item {\bf Declaration of LLM usage}
    \item[] Question: Does the paper describe the usage of LLMs if it is an important, original, or non-standard component of the core methods in this research? Note that if the LLM is used only for writing, editing, or formatting purposes and does \emph{not} impact the core methodology, scientific rigor, or originality of the research, declaration is not required.
    \item[] Answer: \answerNA{}
    \item[] Justification: LLMs are not used as a component of the proposed method.

\end{enumerate}

\end{document}